\documentstyle[preprint,aps,12pt,epsf]{revtex}
\tightenlines
\begin{document}
%\draft command makes pacs numbers print
\draft

%\preprint{\vtop{{\hbox{YITP-99-42}\vskip-0pt\hbox{hep-ph/??????}}}}
\preprint{YITP-99-42}
\title{Non-factorizable contributions in \\ 
nonleptonic weak interactions of $K$ mesons}
\author{K. Terasaki\\ Yukawa Institute for Theoretical Physics,\\
Kyoto University, Kyoto 606-8502, Japan
}
%\date{August 5th, 1999}
\maketitle
\thispagestyle{empty}
\begin{abstract}
$K \rightarrow \pi\pi$, $K_L$-$K_S$ mass difference and 
$K_L \rightarrow \gamma\gamma^{(*)}$ are studied systematically by 
decomposing their amplitude into a sum of factorizable and 
non-factorizable ones. The former is calculated by using the naive 
factorization while the latter is assumed to be dominated by 
dynamical contributions of various hadron states. Non-factorizable 
amplitudes for the $K \rightarrow \pi\pi$ decays are estimated by 
using a hard pion approximation in the infinite momentum frame. 
Long distance non-factorizable contributions to the $K_L$-$K_S$ 
mass difference is dominated by those of pseudo scalar meson poles 
and $\pi\pi$ intermediate states. The amplitude for the 
$K_L \rightarrow \gamma\gamma^{(*)}$ is given by a sum of pole 
contributions of pseudo scalar mesons in the $s$-channel and $K^*$ 
meson in the crossed channel. 

By fitting the results on the $K \rightarrow \pi\pi$, $K_L$-$K_S$ 
mass difference and $K_L \rightarrow \gamma\gamma$ to the 
observations, values of unknown parameters involved are estimated and 
then, by using the resulting values of these parameters, the form 
factor for the Dalitz decays of $K_L$ and their rates are predicted. 
The results are compared with the existing data. 
\end{abstract}

\vskip 0.5cm
%\pacs{PACS number(s): 13.25.-k, 13.25.Es, 12.40.Vv}
%\narrowtext
% body of paper here
%%%%%%%%%%%%%%%%%%%%%%%%%%%%%%%%%%%%%%%%%%%%%%%%%%%%%%%%%%%%%%%%%%%
\newpage
\section{Introduction}

It has been known that short distance contribution is small in the 
$K_L \rightarrow \gamma\gamma$ decay~\cite{Gaillard-Lee} and a 
naively factorized $|\Delta {\bf I}|={1\over 2}$ amplitude for the 
$K \rightarrow \pi\pi$ decays is much smaller than the observed 
one\cite{BBL}. Therefore, in these weak processes, it is expected 
that non-factorizable long distance contribution will play an 
important role. However, in the $K_L$-$K_S$ mass difference 
($\Delta m_K = m_{K_L} - m_{K_S}$), significance of non-factorizable 
contribution is still not clear. Pseudo scalar meson 
poles\cite{Oneda} can give a right size of the mass difference but 
the predicted sign is opposite to the observation (although it is 
sensitive to the $\eta$-$\eta'$ mixing). Contribution of $\pi\pi$ 
intermediate states\cite{Pennington} can occupy about a half of the 
observed mass difference. Short distance contribution has been used 
to test theories within or beyond the standard model by assuming 
explicitly or implicitly dominance of factorizable short distance 
contribution. Therefore, it will be meaningful to study 
systematically these weak processes and check a role of the 
non-factorizable long distance contribution in a overall consistent 
way.  

Our starting point to study nonleptonic weak processes is to 
decompose their amplitude into a sum of {\it factorizable} and 
{\it non-factorizable} ones~\cite{Terasaki-B} and therefore, for 
example, the $|\Delta S|=1$ effective weak Hamiltonian is divided 
into the corresponding parts, 
%%%%%%%%%%%%%%%%%%%%%%%%%%%%%%%%%%%%%%%%%%%%%%%%%%%%%%%%%%%%%%%%%%%%
\begin{equation}
H_w = (H_w)_{\rm FA} + (H_w)_{\rm NF},    
                                                \label{eq:HW-dev}
\end{equation}
%%%%%%%%%%%%%%%%%%%%%%%%%%%%%%%%%%%%%%%%%%%%%%%%%%%%%%%%%%%%%%%%%%%%
where $(H_w)_{\rm FA}$ and $(H_w)_{\rm NF}$ are responsible for
factorizable and non-factorizable amplitudes, respectively. The 
factorizable amplitude is estimated by using the naive 
factorization\cite{Oneda-Wakasa,BSW}. Then, assuming that the 
non-factorizable amplitude is dominated by dynamical contributions of 
various hadrons and using a hard pion approximation in the infinite 
momentum frame (IMF)\cite{hard pion,suppl}, we estimate the 
non-factorizable amplitudes for the $K\rightarrow\pi\pi$ decays. The 
hard pion amplitude will be given by {\it asymptotic} matrix elements 
of $(H_w)_{\rm NF}$ (matrix elements of $(H_w)_{\rm NF}$ taken 
between single hadron states with infinite momentum). 

Before we study explicitly these weak processes, we will investigate
constraints on asymptotic matrix elements of $(H_w)_{\rm NF}$ in the 
next section and find, using a simple quark counting, that the 
asymptotic ground-state-meson matrix elements (matrix elements taken 
between ground-state-meson states with infinite momentum) satisfy the 
$|\Delta {\bf I}|={1\over 2}$ rule. In {Sec.~III}, the 
$K \rightarrow \pi\pi$ decays will be explicitly investigated. 
Since the naively factorized amplitude does not satisfy the 
$|\Delta {\bf I}|={1\over 2}$ rule and its 
$|\Delta {\bf I}|={1\over 2}$ part is much smaller than the observed 
one, the non-factorizable part should dominate the amplitude and 
satisfy the approximate $|\Delta {\bf I}|= {1\over 2}$ rule to 
reproduce the observation. To realize this, in the present approach, 
the asymptotic matrix elements of $(H_w)_{\rm NF}$ should be 
much larger than corresponding ones of $(H_w)_{\rm FA}$ and satisfy 
the $|\Delta {\bf I}|={1\over 2}$ rule as will be shown in {Sec.~II}. 
The $K_L$-$K_S$ mass difference ($\Delta m_K$) will be investigated, 
in {Sec.~IV}, by decomposing it into a sum of {\it short distance} 
and {\it long distance} contributions\cite{Wolfenstein-Hill}, 
%%%%%%%%%%%%%%%%%%%%%%%%%%%%%%%%%%%%%%%%%%%%%%%%%%%%%%%%%%%%%%%%%%%
\begin{equation} 
\Delta m_K = (\Delta m_K)_{\rm SD} + (\Delta m_K)_{\rm LD}.     
                                                        \label{eq:MD}
\end{equation}
%%%%%%%%%%%%%%%%%%%%%%%%%%%%%%%%%%%%%%%%%%%%%%%%%%%%%%%%%%%%%%%%%
Its short distance term, $(\Delta m_K)_{\rm SD}$, is given by 
a matrix element $\langle{K^0|H_{\Delta S=2}|\bar K^0}\rangle$, 
where $H_{\Delta S=2}$ is the $|\Delta S|=2$ effective Hamiltonian 
arising from the so-called box diagrams\cite{Gaillard-Lee}. It will 
be again divided into a sum of factorizable and non-factorizable
parts, 
%%%%%%%%%%%%%%%%%%%%%%%%%%%%%%%%%%%%%%%%%%%%%%%%%%%%%%%%%%%%%%%%%%
\begin{equation}
H_{\Delta S=2} 
= (H_{\Delta S=2})_{\rm FA} + (H_{\Delta S=2})_{\rm NF}.    
                                                   \label{eq:H2-dev}
\end{equation}
%%%%%%%%%%%%%%%%%%%%%%%%%%%%%%%%%%%%%%%%%%%%%%%%%%%
The long distance contribution $(\Delta m_K)_{\rm LD}$ is given by a 
sum of the Born term and the continuum contribution\cite{Itzykson}. 
The former will be dominated by contributions of pseudo-scalar meson 
poles in the present case\cite{Pakvasa}, although contribution of 
vector meson poles has been considered for long
time\cite{Bigi-Sanda}. The latter is dominated by $(\pi\pi)$ 
intermediate states. We will use the result on the $(\pi\pi)$ 
continuum contribution in Ref.\cite{Pennington} as an 
input data. The contribution of pseudo-scalar meson poles will be 
given by asymptotic matrix elements of $H_w$. These matrix elements 
are dominated by non-factorizable ones as will be seen in {Sec.~III}. 
We will study, in {Sec.~V}, two photon decays, 
$K_L \rightarrow \gamma\gamma^{(*)}$, and the Dalitz decays, 
$K_L \rightarrow \gamma\ell^+\ell^-$, where $\gamma^{(*)}$ and $\ell$ 
denote an (off-mass-shell) photon and a lepton, respectively. By
assuming the vector meson dominance (VMD)\cite{VMD}, the amplitude
will be described approximately by two independent matrix elements of 
$(H_w)_{\rm NF}$ taken between pseudo scalar meson states and between 
helicity $\lambda = \pm 1$ vector meson states. Insertion of
constraints on the matrix elements of $(H_w)_{\rm NF}$ (given in 
{Sec.~II}) into the long distance amplitude makes us enable to 
compare our result with experimental data on $K\rightarrow \pi\pi$, 
$\Delta m_K$, $K_L\rightarrow \gamma\gamma$ and 
$K_L\rightarrow\gamma\ell^+\ell^-$ in {Sec.~VI}. A brief summary will 
be given in the final section. 

%%%%%%%%%%%%%%%%%%%%%%%%%%%%%%%%%%%%%%%%%%%%%%%%%%%%%%%%%%%%%%%
\section{Asymptotic Matrix Elements of the Effective Weak
Hamiltonians}

We will describe approximately non-factorizable amplitudes for 
$K\rightarrow \pi\pi$ and $K_L \rightarrow \gamma\gamma^{(*)}$ decays 
as well as the $K^0$-$\bar K^0$ transition using asymptotic matrix 
elements of non-factorizable Hamiltonians, $(H_w)_{\rm NF}$ and 
$(H_{\Delta S=2})_{\rm NF}$, which will be given by color singlet 
sums of colored current products, and then see that the approximate 
$|\Delta {\bf I}|={1\over 2}$ rule in the $K\rightarrow \pi\pi$ 
decays is mainly controlled by the same selection rule in the 
asymptotic ground-state-meson matrix elements of $(H_w)_{\rm NF}$. 
We will see also that 
%%%%%%%%%%%%%%%%%%%%%%%%%%%%%%%%%%%%%%%%%%%%%%%%%%%%%%%%%%%%%
$\langle{K^0|(H_{\Delta S=2})_{\rm NF}|\bar K^0}\rangle$ 
%%%%%%%%%%%%%%%%%%%%%%%%%%%%%%%%%%%%%%%%%%%%%%%%%%%%%%%%%%%%%
is related to 
%%%%%%%%%%%%%%%%%%%%%%%%%%%%%%%%%%%%%%%%%%%%%%%%%%%%%%%%%%%%%%%%%%
$\langle{\pi^0|(H_w^{\Delta I=3/2})_{\rm NF}|\bar K^0}\rangle$ 
%%%%%%%%%%%%%%%%%%%%%%%%%%%%%%%%%%%%%%%%%%%%%%%%%%%%%%%%%%%%%%%%%
and hence the former should vanish if the ground-state-meson matrix
elements of $(H_w)_{\rm NF}$ satisfy the 
$|\Delta {\bf I}|={1\over 2}$ rule, {i.e.}, 
%%%%%%%%%%%%%%%%%%%%%%%%%%%%%%%%%%%%%%%%%%%%%%%%%%%%%%%%%%%%%%%%%%
$\langle{\pi^0|(H_w^{\Delta I=3/2})_{\rm NF}|\bar K^0}\rangle=0$.  
%%%%%%%%%%%%%%%%%%%%%%%%%%%%%%%%%%%%%%%%%%%%%%%%%%%%%%%%%%%%%%%%%

Before we study asymptotic matrix elements of $(H_w)_{\rm NF}$ and 
$(H_{\Delta S=2})_{\rm NF}$, we review briefly these effective 
Hamiltonians. The main part of the $|\Delta S|=1$ effective 
weak Hamiltonian is usually written in the form\cite{HW,SVZ},  
%%%%%%%%%%%%%%%%%%%%%%%%%%%%%%%%%%%%%%%%%%%%%%%%%%%
\begin{equation}
H_w = {G_F \over \sqrt{2}}V_{ud}V_{us}\Bigl\{c_1O_1 + c_2O_2 + 
({\rm penguin})\Bigr\} + h.c.,                        \label{eq:HW}
\end{equation}
%%%%%%%%%%%%%%%%%%%%%%%%%%%%%%%%%%%%%%%%%%%%%%%%%%%
where the four quark operators $O_1$ and $O_2$ are given by products 
of color singlet left-handed currents, 
%%%%%%%%%%%%%%%%%%%%%%%%%%%%%%%%%%%%%%%%%%%%%
\begin{equation}
O_1 = :(\bar us)_{V-A}(\bar du)_{V-A}: \quad {\rm and} \quad 
O_2 = :(\bar uu)_{V-A}(\bar ds)_{V-A}: .          \label{eq:FQO}
\end{equation}
%%%%%%%%%%%%%%%%%%%%%%%%%%%%%%%%%%%%%%%%%%%%%%%
$V_{ij}$ denotes a CKM matrix element\cite{CKM} which is 
taken to be  real since CP invariance is always assumed in this paper. 

When we calculate the factorizable amplitudes for the 
$K \rightarrow \pi\pi$ decays later, we use, as usual, the so-called 
BSW Hamiltonian\cite{BSW,NRSX} 
%%%%%%%%%%%%%%%%%%%%%%%%%%%%%%%%%%%%%%%%%%%%%%%%%%%
\begin{equation}
H_w^{BSW} = {G_F \over \sqrt{2}}V_{ud}V_{us}
\Bigl\{a_1O_1 + a_2O_2 \Bigr\} + h.c. 
                                                   \label{eq:HW-BSW}
\end{equation}
%%%%%%%%%%%%%%%%%%%%%%%%%%%%%%%%%%%%%%%%%%%%%%%%%%%
which can be obtained from Eq.(\ref{eq:HW}) by using the Fierz 
reordering. The operators $O_1$ and $O_2$ in Eq.(\ref{eq:HW-BSW}) 
should be no longer Fierz reordered.  We now replace $(H_w)_{\rm FA}$ 
by $H_w^{BSW}$  as usual. The penguin term has been neglected in 
$H_w^{BSW}$ since its contribution to the naively factorized 
amplitude is expected to be very small\cite{BBL} in  contrast with the 
old expectation~\cite{SVZ}. The coefficients $a_1$  and $a_2$ are given 
by 
%%%%%%%%%%%%%%%%%%%%%%%%%%%%%%%%%%%%%%%%%%%%%%%%%%%%%%%%%%%%%
\begin{equation}
a_1 = c_1 + {c_2 \over N_c}\simeq 1.14,
\quad a_2 = c_2 + {c_1 \over N_c}\simeq -0.209,
                                                \label{eq:coef-BSW}
\end{equation}
%%%%%%%%%%%%%%%%%%%%%%%%%%%%%%%%%%%%%%%%%%%%%%%%%%%%%%
where $N_c$ is the color degree of freedom. Their numerical values 
have been given by using the values of $c_1$ and $c_2$ with the 
leading order QCD corrections\cite{QCD-corr}. 

When $H_w^{BSW}$ is obtained, an extra term which is given by a color 
singlet sum of products of colored currents, 
%%%%%%%%%%%%%%%%%%%%%%%%%%%%%%%%%%%%%%%%%%%%%%%%%%%
\begin{equation}
\tilde H_w = {G_F \over \sqrt{2}}V_{ud}V_{us}
\Bigl\{c_2\tilde O_1 + c_1\tilde O_2 + (\rm{penguin})\Bigr\} + h.c., 
                                                   \label{eq:HW-Non-f}
\end{equation}
%%%%%%%%%%%%%%%%%%%%%%%%%%%%%%%%%%%%%%%%%%%%%%%%%%%
comes out, where 
%%%%%%%%%%%%%%%%%%%%%%%%%%%%%%%%%%%%%%%%%%%%%
\begin{equation}
\tilde O_1 = 2\sum_a:(\bar ut^as)_{V-A}(\bar dt^au)_{V-A}: 
\quad {\rm and} \quad 
\tilde O_2 = 2\sum_a:(\bar ut^au)_{V-A}(\bar dt^as)_{V-A}:         
                                               \label{eq:FQO-extra}
\end{equation}
%%%%%%%%%%%%%%%%%%%%%%%%%%%%%%%%%%%%%%%%%%%%%%%
with the generators $t^a$ of the color $SU_c(N_c)$ symmetry.
Eq.(\ref{eq:HW-Non-f}) can be rewritten in the form, 
%%%%%%%%%%%%%%%%%%%%%%%%%%%%%%%%%%%%%%%%%%%%%%%%%%%
\begin{equation}
\tilde H_w = {G_F \over \sqrt{2}}V_{ud}V_{us}\Bigl\{
\tilde c_-\tilde O_- + \tilde c_+\tilde O_+ + ({\rm penguin})
                                 \Bigr\} + h.c.    \label{eq:HW-pm}
\end{equation}
%%%%%%%%%%%%%%%%%%%%%%%%%%%%%%%%%%%%%%%%%%%%%%%%%%%
where $\tilde O_{\pm} = \tilde O_1 \pm \tilde O_2$. $\tilde O_-$ 
transforms like ${\bf 8_a}$ (not ${\bf 8_s}$) and $\tilde O_+$ 
includes a component transforming like {\bf 27} of the flavor
$SU_f(3)$. They are responsible for the non-factorizable 
$|\Delta {\bf I}|={1\over 2}$ and ${3\over 2}$ amplitudes for the 
$K \rightarrow \pi\pi$ decays, respectively. 

The penguin operators in the SVZ scheme~\cite{SVZ}, 
%%%%%%%%%%%%%%%%%%%%%%%%%%%%%%%%%%%%%%%%%%%%%%%%%%%%%%%%%%%%%%%%%%%
\begin{equation}
O_{5} 
= \sum_a :(\bar dt^as)_{V-A}\sum_q (\bar qt^aq)_{V+A}: 
\quad {\rm and}\quad 
O_{6} 
=  :(\bar ds)_{V-A}\sum_q(\bar qq)_{V+A}: 
\end{equation}
%%%%%%%%%%%%%%%%%%%%%%%%%%%%%%%%%%%%%%%%%%%%%%%%%%%%%%%%%%%%%%%%%%
consist of products of left- and right-handed currents so that 
their contributions to matrix elements taken between two pseudo scalar 
meson states may be enhanced relatively to those of $\tilde O_1$ and 
$\tilde O_2$ which are given by products of left-handed currents. 
Therefore the penguin term cannot necessarily be neglected in 
$\tilde H_w$. Since $O_6$ consists of products of color singlet 
currents, it may be factorizable and included in $H_w^{BSW}$. 
However, the penguin contributions in the factorized amplitudes have 
been small as mentioned before so that $O_5$ which consists of 
colored current products survives in $\tilde H_w$. To realize a 
physical process described in terms of a matrix element, 
$\langle P_2P_3|\tilde H_w|P_1\rangle$, soft gluon(s) have to be 
exchanged between quark(s) and anti-quark(s) belonging to different 
meson states since $\tilde H_w$ is given by a color singlet sum of 
colored current products as seen the above. Therefore, 
$\langle P_2P_3|\tilde H_w|P_1\rangle$ is not factorizable, i.e., 
$(H_w)_{\rm NF}$ is given by $\tilde H_w$. In this way, the 
$|\Delta S|=1$ effective weak Hamiltonian is now given in the form, 
%%%%%%%%%%%%%%%%%%%%%%%%%%%%%%%%%%%%%%%%%%%%%%%%%%%
\begin{equation}
H_w \rightarrow  H_w^{BSW} + \tilde H_w.            \label{eq:HW-tot}
\end{equation}
%%%%%%%%%%%%%%%%%%%%%%%%%%%%%%%%%%%%%%%%%%%%%%%%%%%

In the same way, the $\Delta S=2$ effective Hamiltonian, 
%%%%%%%%%%%%%%%%%%%%%%%%%%%%%%%%%%%%%%%%%%%%%%%%%%%
\begin{equation}
H_{\Delta S=2}=c_{\Delta S=2}O_{\Delta S=2}\,\, + \,\, h.c \quad 
{\rm with}  
\quad   O_{\Delta S=2} = :(\bar ds)_{V-A}(\bar ds)_{V-A}: ,
\label{eq:H(S=2)}
\end{equation}
%%%%%%%%%%%%%%%%%%%%%%%%%%%%%%%%%%%%%%%%%%%%%%%%%%%
is again replaced by a sum of the Fierz reordered BSW-like 
Hamiltonian, $H_{\Delta S=2}^{BSW}$, and an extra term, 
$\tilde H_{\Delta S=2}$, i.e., 
%%%%%%%%%%%%%%%%%%%%%%%%%%%%%%%%%%%%%%%%%%%%%%%%%%%%%%%%%%%%%%%% 
\begin{equation}
H_{\Delta S=2} 
\rightarrow H_{\Delta S=2}^{BSW} + \tilde H_{\Delta S=2}, 
                                             \label{eq:H(S=2)-tot}
\end{equation}
%%%%%%%%%%%%%%%%%%%%%%%%%%%%%%%%%%%%%%%%%%%%%%%%%%%
where $H_{\Delta S=2}^{BSW}$ is obtained by applying the Fierz
reordering to the above $H_{\Delta S=2}$. The extra term is again
given by a color singlet sum of products of colored currents, 
%%%%%%%%%%%%%%%%%%%%%%%%%%%%%%%%%%%%%%%%%%%%%%%%%%%
\begin{equation}
\tilde H_{\Delta S=2}
=c_{\Delta S=2}\tilde O_{\Delta S=2}\,\,+\,\,h.c. 
\quad {\rm with}\quad \tilde O_{\Delta S=2} 
             = 2\sum_a:(\bar dt^as)_{V-A}(\bar dt^as)_{V-A}:. 
                                              \label{eq:H(S=2)-extra}
\end{equation}
%%%%%%%%%%%%%%%%%%%%%%%%%%%%%%%%%%%%%%%%%%%%%%%%%%%

Now we study constraints on matrix elements of non-factorizable 
Hamiltonians\cite{quark-counting,TBD-charm-fact}. The 
non-factorizable four-quark operators $\tilde O_\pm$ and 
$\tilde O_{\Delta S=2}$ can be expanded into a sum of products of 
(a) two creation operators to the left and two annihilation operators 
to the right, (b) three creation operators to the left and one 
annihilation operator to the right, (c) one creation operator to the 
left and three annihilation operators to the right, and (d) all 
(four) creation operators or annihilation operators of quarks and 
anti-quarks. We associate (a)$-$(d) with quark-line diagrams 
describing different types of matrix elements of $\tilde O_\pm$ and 
$\tilde O_{\Delta S=2}$. For (a), we utilize the two creation and 
annihilation operators to create and annihilate, respectively, the 
quarks and anti-quarks belonging to the meson states 
$\langle \{q\bar q\}|$ and $|\{q\bar q\}\rangle$ in the asymptotic 
matrix elements of the four-quark operators, $\tilde O_\pm$ and 
$\tilde O_{\Delta S=2}$. For (b) and (c), we need to add a spectator 
quark or anti-quark to reach {\it physical} processes, for example, 
%%%%%%%%%%%%%%%%%%%%%%%%%%%%%%%%%%%%%%%%%%%%%%%%%%%%%%%%%%%%%%%%%%%%
$\langle {\{qq\bar q\bar q\}|\tilde O_\pm|\{q\bar q\}}\rangle$ and 
$\langle {\{q\bar q\}|\tilde O_\pm|\{qq\bar q\bar q\}}\rangle$.  
%%%%%%%%%%%%%%%%%%%%%%%%%%%%%%%%%%%%%%%%%%%%%%%%%%%%%%%%%%%%%%%%%%%%
Here $\{qq\bar q\bar q\}$ denotes a four-quark meson\cite{Jaffe}. 

While we count all possible connected quark-line diagrams, we forget 
color degree of freedom of quarks since they will be compensated by a 
deep sea of soft gluons carried by light mesons. However, we have to
be careful with the order of the quark(s) and anti-quark(s) in 
$\tilde O_\pm$ and $\tilde O_{\Delta S=2}$ since symmetry (or 
antisymmetry) property of wave functions of meson states under 
exchanges of quark and anti-quark plays an important role when 
asymptotic matrix elements of $\tilde O_\pm$ and 
$\tilde O_{\Delta S=2}$ are considered. Noting that the wave function 
of the ground-state $\{q\bar q\}_0$ meson is antisymmetric under 
exchange of its quark and anti-quark\cite{CLOSE}, we
obtain~\cite{TBD}, 
%%%%%%%%%%%%%%%%%%%%%%%%%%%%%%%%%%%%%%%%%%%%%%%%%%%%%%%%
\begin{equation}
\langle{\{q\bar q\}_0|\tilde O_+|\{q\bar q\}_0}\rangle = 0, 
                                                \label{eq:SUM-G}
\end{equation}
%%%%%%%%%%%%%%%%%%%%%%%%%%%%%%%%%%%%%%%%%%%%%%%%%%%%%%%%
which implies that the asymptotic ground-state-meson matrix elements 
of $\tilde H_w$ satisfy the $|\Delta{\bf I}|={1\over 2}$ rule since
the penguin term satisfies it always. We will neglect contributions 
of excited $\{q\bar q\}$ meson states as will be discussed in the 
next section. The same quark counting leads directly 
to\cite{Terasaki-FF,Delta m_K-asymp} 
%%%%%%%%%%%%%%%%%%%%%%%%%%%%%%%%%%%%%%%%%%%%%%%%%%%%%%%%%%%%%
\begin{equation}
\langle{K^0|\tilde O_{\Delta S=2}|\bar K^0}\rangle = 0
                                              \label{eq:SUM-G_{S=2}}
\end{equation}
%%%%%%%%%%%%%%%%%%%%%%%%%%%%%%%%%%%%%%%%%%%%%%%%%%%%%%%%%%%%%
which is compatible with Eq.(\ref{eq:SUM-G}). 

According to Ref.\cite{Jaffe}, four quark  $\{qq\bar q\bar q\}$ 
mesons are classified into the following four types, 
%%%%%%%%%%%%%%%%%%%%%%%%%%%%%%%%%%%%%%%%%%%%%%%%%%%%%%%%%%%%%%%%%%%%
$\{qq\bar q\bar q\} = [qq][\bar q\bar q] \oplus (qq)(\bar q\bar q) 
\oplus \{[qq](\bar q\bar q) \pm (qq)[\bar q\bar q]\}$,  
%%%%%%%%%%%%%%%%%%%%%%%%%%%%%%%%%%%%%%%%%%%%%%%%%%%%%%%%%%%%%%%%%%%%
where () and [] denote symmetry and antisymmetry, respectively, under 
the exchange of flavors between them. The first two can have 
$J^{P}=0^{+}$ but the last two have $J^{P}=1^{+}$. Each 
multiplet classified above is again classified into two different
classes according to different combinations of color degree of 
freedom. However, they can mix with each other. The masses of 
four-quark mesons with the mixing have been predicted by using the 
bag model. The members of the heavier class have been expected to 
play an important role in charm decays\cite{TBD-charm-fact} since 
their predicted masses are close to the parent charm meson masses 
while four-quark meson contribution to $K$ decays will be small 
since their masses are considerably larger than the kaon mass $m_K$. 
However, in the $K^+\rightarrow\pi^+\pi^0$ decay, their contributions 
are not necessarily negligible since the other contributions are 
suppressed because of the above $|\Delta{\bf I}|={1\over 2}$ rule in 
the asymptotic ground-state-meson matrix elements of $\tilde H_w$,
etc. We here take contributions of the $[qq][\bar q\bar q]$ and 
$(qq)(\bar q\bar q)$ with $J^{P(C)}=0^{+(+)}$ belonging to the 
lower mass class, although contributions of heavier ones may not 
be negligible for more precise discussions. 

Using the same prescription as the above, we can obtain the following 
constraints on asymptotic matrix elements of $\tilde H_w$ between 
$\{q\bar q\}_0$ and $\{qq\bar q\bar q\}$ meson states~\cite{TBD}, 
%%%%%%%%%%%%%%%%%%%%%%%%%%%%%%%%%%%%%%%%%%%%%%%%%%%%%%%%
\begin{eqnarray}
&&\langle{[qq][\bar q\bar q]|\tilde O_+|\{q\bar q\}_0}\rangle
= \langle{\{q\bar q\}_0|\tilde O_+|[qq][\bar q\bar q]}\rangle = 0, 
                                             \label{eq:SUM-anti}\\
&&\langle{(qq)(\bar q\bar q)|\tilde O_-|\{q\bar q\}_0}\rangle 
= \langle{\{q\bar q\}_0|\tilde O_-|(qq)(\bar q\bar q)}\rangle =0. 
                                                 \label{eq:SUM-sym}
\end{eqnarray}
%%%%%%%%%%%%%%%%%%%%%%%%%%%%%%%%%%%%%%%%%%%
The above equations imply that asymptotic matrix elements of 
$\tilde H_w$ between $\{q\bar q\}_0$ and $[qq][\bar q\bar q]$ meson 
states satisfy the $|\Delta{\bf I}|={1\over 2}$ rule while the matrix 
elements between $\{q\bar q\}_0$ and $(qq)(\bar q\bar q)$ can violate 
the rule. Therefore the $|\Delta{\bf I}|={1\over 2}$ rule violating 
non-factorizable amplitude for the $K\rightarrow \pi\pi$ decays can 
be supplied through the $(qq)(\bar q\bar q)$ meson pole amplitudes 
which can interfere destructively with the too big 
$|\Delta{\bf I}|={3\over 2}$ part of the factorized amplitudes in 
Table~I. 

The ground-state-meson matrix elements of $O_5$ is also treated in the 
same way. Since the flavor $SU_f(3)$ singlet $q\bar q$ pairs in the
$s$-channel can couple to a glue(-ball) state and induce a matrix
element $\langle{g_0|O_5|\bar K^0}\rangle$, where $g_0$ denotes 
a glue-ball, we treat separately with such contributions when we 
consider matrix elements $\langle{P|O_5|\bar K^0}\rangle$ with 
$P=\pi^0$, $\eta$, $\eta'$, $\ldots\,$ In this way, we obtain
%%%%%%%%%%%%%%%%%%%%%%%%%%%%%%%%%%%%%%%%%%%%%%%%%%%%%%%%%%%%%%%%%%%%
\begin{equation}
\langle{\pi^-|O_5|K^-}\rangle
=-\sqrt{2}\langle{\pi^0|O_5|\bar K^0}\rangle
=\sqrt{2}\langle{\eta_0|O_5|\bar K^0}\rangle
=\langle{\eta_s|O_5|\bar K^0}\rangle, \label{eq:SUM-O5}
\end{equation}
%%%%%%%%%%%%%%%%%%%%%%%%%%%%%%%%%%%%%%%%%%%%%%%%%%%%%%%%%%%%%%%%%%%%
where 
%%%%%%%%%%%%%%%%%%%%%%%%%%%%%%%%%%%%%%%%%%%%%%%%%%%%%%%%%%%%%%%%%%%%%
$\eta_{0} \sim (u\bar u + d\bar d)/\sqrt{2}$ and 
$\eta_s \sim (s\bar s)$ 
%%%%%%%%%%%%%%%%%%%%%%%%%%%%%%%%%%%%%%%%%%%%%%%%%%%%%%%%%%%%%%%%%%%%%%
are components of iso-singlet pseudo scalar mesons, $\eta$ and 
$\eta'$, 
%%%%%%%%%%%%%%%%%%%%%%%%%%%%%%%%%%%%%%%%%%%%%%%%%%%%%%%%%%%%%%%%%%%%%
\begin{equation}
\left\{ 
\begin{array}{l}
\eta \,\,= a_\eta^0\,\eta_0\,\,+\,a_\eta^s\,\eta_s \,
= \Bigl(\sqrt{1\over 3}{\rm cos}{\theta_P} 
            \, - \sqrt{2\over 3}{\rm sin}{\theta_P}\Bigr)\eta_0 
     - \Bigl(\quad\sqrt{2\over 3}{\rm cos}{\theta_P} 
             + \sqrt{1\over 3}{\rm sin}{\theta_P}\Bigr)\eta_s, 
                                                    \label{eq:eta}\\
\eta'= a_{\eta'}^0\,\eta_0\,+\,a_{\eta'}^s\,\eta_s
= \Bigl(\sqrt{1\over 3}{\rm sin}{\theta_P} 
            \,  + \sqrt{2\over 3}{\rm cos}{\theta_P}\Bigr)\eta_0 
    + \Bigl(-\,\sqrt{2\over 3}{\rm sin}{\theta_P} 
              + \sqrt{1\over 3}{\rm cos}{\theta_P}\Bigr)\eta_s. 
                                                % \label{eq:eta'} 
\end{array} 
\right.
\end{equation}
%%%%%%%%%%%%%%%%%%%%%%%%%%%%%%%%%%%%%%%%%%%%%%%%%%%%%%%%%%%%%%%%%%%%%
The mixing angle is usually taken to be 
$\theta_P \simeq -20^\circ$\cite{PDG}. 

Now we are ready to parameterize the asymptotic ground-state-meson 
matrix elements of $\tilde H_w$. To reproduce the observed 
$|\Delta{\bf I}|={1\over 2}$ rule in the $K\rightarrow \pi\pi$ 
decays, we need the $|\Delta{\bf I}|={1\over 2}$ rule for the 
ground-state-meson matrix elements of $\tilde H_{w}$ with a 
sufficient precision. It is all right if one accepts the above quark 
counting. (If not, one should assume the $|\Delta{\bf I}|={1\over 2}$ 
rule for the ground-state-meson matrix elements of $\tilde H_{w}$.) 
Anyway, neglecting seemingly small (or {\it zero} in the above quark 
counting) $|\Delta{\bf I}|={3\over 2}$ contributions, we parameterize 
the (asymptotic) ground-state-meson matrix elements of $\tilde H_{w}$ 
as follows,  
%%%%%%%%%%%%%%%%%%%%%%%%%%%%%%%%%%%%%%%%%%%%%%%%%%%%%%%%%%%%%%%%%%%%
\hfil\break 
(A) helicity $\lambda=0$ matrix elements:
%%%%%%%%%%%%%%%%%%%%%%%%%%%%%%%%%%%%%%%%%%%%%%%%%%%%%%%%%%%%%%%%%
\begin{equation}
\left\{ 
\begin{array}{l}
\langle{\pi^-|\tilde H_{w}|K^-}\rangle 
= \qquad\, (1 + r_{0})H_{0},   \quad\,
\langle{\pi^0\,|\tilde H_{w}\,|\bar K^0\,}\rangle 
= -\sqrt{1\over 2}(1 + \tilde r_{0})\tilde H_{0},   \\
\langle{\eta_{0}\,\,|\tilde H_{w}|\bar K^0\,}\rangle 
= -\sqrt{1\over 2}(1 - \tilde r_{0})\tilde H_{0},   \quad\,
\langle{\eta_{s}\,\,|\tilde H_{w}\,|\bar K^0\,}\rangle 
=\qquad\qquad\,\,\,  \tilde r_{0}\,\tilde H_{0},   
\label{eq:para-0}
\end{array} 
\right.
\end{equation}
%%%%%%%%%%%%%%%%%%%%%%%%%%%%%%%%%%%%%%%%%%%%%%%%%%%%%%%%%%%%%%%%%
(B) helicity $|\lambda|=1$ matrix elements:
%%%%%%%%%%%%%%%%%%%%%%%%%%%%%%%%%%%%%%%%%%%%%%%%%%%%%%%%%%%%%%%%%
\begin{equation}
\left\{ 
\begin{array}{l}
\langle{\rho^0\,|\tilde H_{w}|\bar K^{*0}}\rangle_{1}
= -\sqrt{1\over 2}(1 + \tilde r_{1})\tilde H_{1},   \quad
\langle{\,\omega\,|\tilde H_{w}|\bar K^{*0}}\rangle_{1} 
= -\sqrt{1\over 2}(1 - \tilde r_{1})\tilde H_{1},    \\
\langle{\,\phi\,\,|\tilde H_{w}|\bar K^{*0}}\rangle_{1} 
= \qquad\qquad\,\, \tilde r_{1}\,\,\tilde H_{1}.    
                                         \label{eq:para-1}
\end{array} 
\right.
\end{equation}
%%%%%%%%%%%%%%%%%%%%%%%%%%%%%%%%%%%%%%%%%%%%%%%%%%%%%%%%%%%%%%%%%%%
The $\omega$-$\phi$ mixing has been assumed to be ideal. The
parameters $\tilde r_0$ and $\tilde r_1$ denote contributions of the 
penguin relative to $\tilde O_-$ in the helicity $\lambda=0$ and 
$|\lambda|= 1$ matrix elements of $\tilde H_w$, respectively. 
$\tilde H_0$ and $\tilde H_1$ provide their normalizations. In (B), 
$\tilde r_1$ will be neglected hereafter since it is expected to be 
small because of the small coefficient of the penguin and because of 
a helicity consideration. 

Constraints on asymptotic matrix elements of $\tilde H_w$ from 
Eqs.(\ref{eq:SUM-anti}) and (\ref{eq:SUM-sym}) including four-quark 
meson states and their parameterizations will be given in 
{Appendix A}.

%%%%%%%%%%%%%%%%%%%%%%%%%%%%%%%%%%%%%%%%%%%%%%%%%%%%%%%%%%%%%%%
\section{Two pion decays of $K$ mesons}

Now we study two pion decays of $K$ mesons. As mentioned before, an 
amplitude for $K\rightarrow\pi\pi$ decay is given by a sum of 
{\it factorizable} and {\it non-factorizable} ones. The former 
is estimated by using the naive factorization below while the latter 
is assumed to be dominated by dynamical contributions of various 
hadron states and will be estimated later by using a hard pion 
approximation in the IMF. 

The factorizable amplitudes for the $K \rightarrow \pi\pi$ decays 
are estimated by using the naive factorization in the BSW 
scheme\cite{BSW}. As an example, we consider the amplitude for the 
$K^+ \rightarrow \pi^+\pi^0$ decay. It is given by 
%%%%%%%%%%%%%%%%%%%%%%%%%%%%%%%%%%%%%%%%%%%%%%%%%%%%%%%%%%%%%%%%%%%
\begin{eqnarray}
&& M_{\rm FA}(K^+ \rightarrow \pi^0\pi^+) 
= \langle {\pi^+(q)\pi^0(p')|H_w^{BSW}|K^+(p)}\rangle  
\nonumber\\
&&\qquad = {G_F \over \sqrt{2}}V_{us}V_{ud}
\Bigl\{ 
a_1\langle \pi^+(q)|(\bar ud)_{V-A}|0\rangle 
\langle \pi^0(p')|(\bar su)_{V-A}|K^+(p) \rangle         \nonumber\\
&&\hspace{3cm} + a_2\langle \pi^0(p')|(\bar uu)_{V-A}|0\rangle
\langle \pi^+(q)|(\bar sd)_{V-A}|K^+(p) \rangle
\Bigr\}.
                                                    \label{eq:FACT}
\end{eqnarray}
%%%%%%%%%%%%%%%%%%%%%%%%%%%%%%%%%%%%%%%%%%%%%%%%%%%%%%%%%%%%%%%%%%%%
Factorizable amplitudes for the other $K \rightarrow \pi\pi$ decays 
also can be calculated in the same way. To evaluate these amplitudes, 
we use the following parameterization of matrix elements 
of currents, 
%%%%%%%%%%%%%%%%%%%%%%%%%%%%%%%%%%%%%%%%%%%%%%%
\begin{equation}
\left\{ 
\begin{array}{l}
\langle \pi(q)|A_\mu |0 \rangle 
                         = -if_{\pi}q_\mu, \quad {\rm etc.}, \\
\langle \pi(p')|V_\mu |K(p)\rangle 
= (p+p')_\mu f_+^{(\pi K)}(q^2) 
              + q_\mu f_-^{(\pi K)}(q^2), \quad {\rm etc.}, 
\end{array} 
\right.
\end{equation}
%%%%%%%%%%%%%%%%%%%%%%%%%%%%%%%%%%%%%%%%%%%%%%%%%%%%%%%%%%%%%%%%%%%%
where $q = p - p'$. Using these expressions of current matrix 
elements, we obtain the factorized amplitudes listed in Table~I, 
where terms proportional to $f_-(q^2)$ have been neglected since 
their coefficients are small in the spectator decays and, in possible 
annihilation decays, they 
%%%%%%%%%%%%%%%%%%%%%%%%%%%%%%%%%%%%%%%%%%%%%%%%%%%%%%%%%%
%\newpage
%\vspace{0.5cm}
%\begin{table}[t]
\begin{center}
\begin{quote}
{Table~I. Naively factorized amplitudes for the $K\rightarrow\pi\pi$ 
decays, where terms proportional to $f_-$ are neglected.}
\end{quote}
\vspace{0.5cm}

\begin{tabular}
{l l}
\hline\hline
$\quad\,\,${\rm Decay}
&\hskip 3.5cm {$\quad M_{\rm FA}\,$}
\\
\hline 
\vspace{-3mm}
\\
$K_S\, \rightarrow \pi^+\pi^-$
& $\quad\quad iV_{ud}V_{us}
(\displaystyle{G_F\over\sqrt{2}})
          \sqrt{2}a_1f_\pi(m_K^2 - m_\pi^2)f_+^{\pi K}(m_\pi^2)$
\vspace{2mm}\\
%\hline
$K_S\, \rightarrow \pi^0\,\pi^0$
& $\quad-\,iV_{ud}V_{us}
(\displaystyle{G_F\over\sqrt{2}})
                  a_2f_\pi(m_K^2 - m_\pi^2)f_+^{\pi K}(m_\pi^2)$
\vspace{2mm}\\
%\hline
$K^+ \rightarrow \pi^+\pi^0$
& $\quad\quad iV_{ud}V_{us}(\displaystyle{G_F\over{2}})
(a_1+a_2)f_\pi(m_K^2 - m_\pi^2)f_+^{\pi K}(m_\pi^2)$ 
\vspace{2mm}\\
\hline\hline
\end{tabular}

%\end{table}
\end{center}
\vspace{1.0cm}
%%%%%%%%%%%%%%%%%%%%%%%%%%%%%%%%%%%%%%%%%%%%%%%%%%%%%%%%%%%%%%%%%%
are proportional to the small coefficient 
$a_2$. The naively factorized penguin contribution has also been 
neglected as discussed in Sec. II. If the values of $a_1$ and $a_2$ 
with the leading order QCD corrections\cite{BBL} are taken, it will 
be seen, since $|a_1| \gg |a_2|$, that the factorized amplitude for 
the $K^0\rightarrow \pi^0\pi^0$ decay which is described by the color 
mismatched diagram, $\bar s \rightarrow \bar d\,+\,(u\bar u)_1$, is 
proportional to $a_2$ and therefore is much smaller (the color 
suppression) than those for the spectator decays, where $(u\bar u)_1$
denotes a color singlet pair of $u$ and $\bar u$. It is also seen 
that the factorized amplitude for the $K^+\rightarrow\pi^+\pi^0$ 
decay is considerably larger than the observed one and that the size 
of the $|\Delta{\bf I}|={1\over 2}$ amplitude is not much larger than 
the $|\Delta{\bf I}|={3\over 2}$ part. Therefore it is hard to 
reproduce the well-known approximate $|\Delta{\bf I}|={1\over 2}$ 
rule by the naively factorized amplitudes for the 
$K \rightarrow \pi\pi$ decays. 

Next we study non-factorizable amplitudes for these decays using a 
hard pion approximation in the IMF\cite{hard pion,suppl}, {i.e.}, we 
evaluate the amplitudes at a slightly unphysical point 
${\bf q}\rightarrow 0$ in the IMF (${\bf p}\rightarrow \infty$) in 
which $q_0 \rightarrow O(|{\bf p}|^{-1})$ and hence $(p\cdot q)$ 
is still finite. Then the hard pion amplitude as the non-factorizable 
one is written in the form, 
%%%%%%%%%%%%%%%%%%%%%%%%%%%%%%%%%%%%%%%%%%%%%%%%%%%%%%%%%%%%%%%%%%
\begin{equation}
M_{\rm NF}(K\rightarrow \pi_1\pi_2) 
\simeq M_{\rm ETC}(K\rightarrow \pi_1\pi_2) 
+ M_{\rm surf}(K\rightarrow \pi_1\pi_2),            \label{eq:hard pion}
\end{equation}
%%%%%%%%%%%%%%%%%%%%%%%%%%%%%%%%%%%%%%%%%%%%%%%%%%%%%%%%%%%%%%%%% 
where $M_{\rm ETC}$ and $M_{\rm surf}$ are given by 
%%%%%%%%%%%%%%%%%%%%%%%%%%%%%%%%%%%%%%%%%%%%%%%%%%%%%%%%%%%%%%
\begin{equation}
M_{\rm ETC}(K\rightarrow \pi_1\pi_2)
= {i \over \sqrt{2}f_{\pi}}
     \langle{\pi_2|[V_{\bar \pi_1}, \tilde H_w]|K}\rangle 
                      + (\pi_2 \leftrightarrow \pi_1)   \label{eq:ETC}
\end{equation}
%%%%%%%%%%%%%%%%%%%%%%%%%%%%%%%%%%%%%%%%%%%%%%%%%%%%%%
and
%%%%%%%%%%%%%%%%%%%%%%%%%%%%%%%%%%%%%%%%%%%%%%%%%%%%%%%%%%%%%%%%  
\begin{eqnarray} 
&&M_{\rm surf}(K\rightarrow \pi_1\pi_2)
= {i \over \sqrt{2}f_{\pi}}
\Biggl\{\sum_n\Bigl({m_{\pi}^2 - m_{K}^2 
                                 \over m_n^2 - m_{K}^2}\Bigr)
  \langle{\pi_2|A_{\bar \pi_1}|n}\rangle
                         \langle{n|\tilde H_w|K}\rangle  \nonumber\\
&&\hspace{5cm} + \sum_l\Bigl({m_{\pi}^2 - m_{K}^2 
                              \over m_l^2 - m_{\pi}^2}\Bigr)
\langle{\pi_2|\tilde H_w|l}\rangle
                  \langle{l|A_{\bar \pi_1}|K}\rangle\Biggr\} 
+ (\pi_2 \leftrightarrow \pi_1),  
                                                    \label{eq:SURF}
\end{eqnarray}
%%%%%%%%%%%%%%%%%%%%%%%%%%%%%%%%%%%%%%%%%%%%%%%%%%%%%%%%
respectively, where $[V_\pi + A_\pi, \tilde H_w]=0$ has been 
used~\cite{Donaghue} since $\tilde O_{\pm}$ consist of left-handed 
currents and the right-handed currents in the penguin term is of 
flavor singlet. The equal-time commutator term, $M_{\rm ETC}$, has 
the same form as the one in the old soft pion 
approximation~\cite{soft-pion} but now has to be evaluated in the IMF. 
The surface term, $M_{\rm surf}$, is given by a divergent of matrix 
element of $T$-product of axial vector current and $\tilde H_w$ taken
between $\langle\pi|$ and $|K\rangle$. However, in contrast with 
the soft pion approximation, contributions of single meson 
intermediate states can now survive, when complete sets 
of energy eigen states are inserted between these two operators. 
(See, for more details, Refs.\cite{hard pion} and \cite{suppl}.) 
Therefore, $M_{\rm surf}$ is given by a sum of all possible pole 
amplitudes, i.e., $n$ and $l$ in Eq.(\ref{eq:SURF}) run over all 
possible single meson states, not only ordinary $\{q\bar q\}$, but 
also hybrid $\{q\bar qg\}$, four-quark $\{qq\bar q\bar q\}$, 
glue-balls, etc. However, values of wave functions of orbitally
excited $\{q\bar q\}_{L \neq 0}$ states at the origin are expected to 
vanish in the non-relativistic quark model and, more generally, wave 
function overlappings between the ground-state $\{q\bar q\}_0$ and 
excited-state-meson states are expected to be small so that 
excited-state-meson contributions are not very important except for 
the non-factorizable $K^+\rightarrow \pi^+\pi^0$ amplitude in which 
the ground-state-meson contributions can be strongly suppressed
because of the $|\Delta{\bf I}|={1\over 2}$ rule in the asymptotic 
ground-state-meson matrix elements of $\tilde H_w$ as seen in 
Sec.~II. Asymptotic matrix elements of isospin $V_\pi$ and its axial 
counterpart $A_\pi$ involved in the amplitudes can be well 
parameterized by using (asymptotic) $SU_f(3)$ symmetry. Therefore 
the hard pion amplitude in Eq.(\ref{eq:hard pion}) with 
Eqs.(\ref{eq:ETC}) and (\ref{eq:SURF}) as the non-factorizable long 
distance contribution is approximately described in terms of 
asymptotic ground-state-meson matrix elements of $\tilde H_w$. 

Amplitudes for dynamical hadronic processes, in general, can be 
described in the form, 
%%%%%%%%%%%%%%%%%%%%%%%%%%%%%%%%%%%%%%%%%%%%%%%%%%%%%%%%%%%%%%%%
\begin{equation}
({\rm continuum \,\,\, contribution})\,\,\,\,
+ \,\,\,\,({\rm  Born\,\,\, term}).  
\end{equation}
%%%%%%%%%%%%%%%%%%%%%%%%%%%%%%%%%%%%%%%%%%%%%%%%%%%%%%%%%%%%%%%
In the present case, $M_{\rm surf}$ is given by a sum of pole 
amplitudes so that $M_{\rm ETC}$ corresponds to the continuum 
contribution\cite{MATHUR} which can develop a phase relative to the 
Born term. Therefore, using isospin eigen amplitudes,
$M_{ETC}^{(I)}$'s, and their phases, $\delta_I$'s, we here 
parameterize the ETC terms as 
%%%%%%%%%%%%%%%%%%%%%%%%%%%%%%%%%%%%%%%%%%%%%%
\begin{equation}
\left\{ 
\begin{array}{l}
M_{\rm ETC}(K_S^0\, \rightarrow \pi^+\pi^-) 
= \quad\,\,\,\displaystyle{2 \over 3}\,\,\,\,
     M_{\rm ETC}^{(2)}(K\rightarrow \pi\pi) e^{i\delta_{2}}\, 
+\,\,\, \displaystyle{1 \over 3}\,\,
     M_{\rm ETC}^{(0)}(K\rightarrow \pi\pi) e^{i\delta_{0}},
                                                \vspace{2mm} \\
M_{\rm ETC}(K_S^0\, \rightarrow \pi^0\,\pi^0\,) \,
= -\displaystyle{2\sqrt{2} \over 3}
     M_{\rm ETC}^{(2)}(K\rightarrow \pi\pi) e^{i\delta_{2}}\, 
+  \displaystyle{\sqrt{1\over 2}}
     M_{\rm ETC}^{(0)}(K\rightarrow \pi\pi) e^{i\delta_{0}}, 
                                                \vspace{2mm} \\
M_{\rm ETC}(K^+ \rightarrow  \pi^+\pi^0\,) \,
= \qquad\,\,\,\, M_{\rm ETC}^{(2)}(K\rightarrow \pi\pi)
                                     e^{i\delta_{2}}, 
\end{array} 
\right.
\end{equation}
%%%%%%%%%%%%%%%%%%%%%%%%%%%%%%%%%%%%%%%%%%%%%%%
since the $S$-wave $\pi\pi$ final states can have isospin $I=0$ and 
2. The so-called final state interactions are given by dynamics of
hadrons and therefore now they are included in the non-factorizable 
long distance amplitudes. 

By neglecting small contributions of excited states and 
$|\Delta{\bf I}|={3\over 2}$ asymptotic ground-state-meson matrix
elements of $\tilde H_w$ which vanish in the quark counting as seen 
before, the non-factorizable amplitudes for the $K\rightarrow \pi\pi$ 
decays can be approximately given as follows, 
%%%%%%%%%%%%%%%%%%%%%%%%%%%%%%%%%%%%%%%%%%%%%%%%%%%%%%%%%%
\begin{equation}
\left\{ 
\begin{array}{l}
M_{\rm NF}(K_S^0\, \rightarrow \pi^+\pi^-)
\simeq -\displaystyle{i \over f_\pi}
                \langle{\pi^+|\tilde H_w|K^+}\rangle e^{i\delta_0},   
                                                     \vspace{2mm} \\
M_{\rm NF}(K_S^0\, \rightarrow \pi^0\,\pi^0\,)
\simeq -\displaystyle{\sqrt{1\over 2}}
      M_{\rm NF}(K_S \rightarrow \pi^+\pi^-),\vspace{2mm}  \\
M_{\rm NF}(K^+ \rightarrow \pi^+\pi^0\,) \simeq \,\,0, 
                                                   \label{eq:non-f-0}
\end{array} 
\right.
\end{equation}
%%%%%%%%%%%%%%%%%%%%%%%%%%%%%%%%%%%%%%%%%%%%%%%%%%%%%%%%%%%%%%%%%%
where $K^*$ meson pole contributions in the $u$-channel has been
neglected\cite{Pakvasa} although they have been accepted for long
time\cite{Marshak}. It is seen that the non-factorizable amplitudes
for $K\rightarrow \pi\pi$ decays have been described approximately in 
terms of the asymptotic ground-state-meson matrix elements of $\tilde
H_w$ and the iso-scalar $S$-wave $\pi\pi$ phase shift $\delta_0$ and 
that they satisfy the $|\Delta {\bf I}|={1\over 2}$ rule. More
accurate amplitudes involving excited (four-quark) meson 
contributions will be given in {Appendix B}. In the last equation of 
Eq.(\ref{eq:non-f-0}), the right-hand-side is vanishing since
the four-quark meson contributions have been neglected. As seen in 
{Appendix B}, the $|\Delta{\bf I}|={3\over 2}$ part of the
non-factorizable amplitude can be supplied through contributions of 
four-quark meson  poles since matrix elements of $\tilde H_w$ taken 
between four-quark $(qq)(\bar q\bar q)$ and the ground-state 
$\{q\bar q\}_0$ meson states can violate the 
$|\Delta{\bf I}|={1\over 2}$ rule as seen in {Appendix A}. 
They can interfere destructively with the too big 
$|\Delta{\bf I}|={3\over 2}$ part of the factorized amplitudes in 
Table~I. 

%%%%%%%%%%%%%%%%%%%%%%%%%%%%%%%%%%%%%%%%%%%%%%%%%%%%%%%%%%%%%%%
\section{$K_L$-$K_S$ mass difference}

Now we study the $K_L$-$K_S$ mass difference, $\Delta m_K$, by 
decomposing it into a sum of {\it short distance} and 
{\it long distance} contributions, 
%%%%%%%%%%%%%%%%%%%%%%%%%%%%%%%%%%%%%%%%%%%%%%%%%%%%%%%% 
\begin{equation}
\Delta m_K = (\Delta m_K)_{\rm SD} + (\Delta m_K)_{\rm LD},  
                                              \label{eq:Delta-m_K}
\end{equation}
%%%%%%%%%%%%%%%%%%%%%%%%%%%%%%%%%%%%%%%%%%%%%%%%%%%%%%%%%%%%%%%
as usual\cite{Delta-m_K}. The short distance contribution 
$(\Delta m_K)_{\rm SD}$ is given by the matrix element of the 
$\Delta S=2$ box operator\cite{Gaillard-Lee} taken between 
$\langle{K^0}|$ and $|{\bar K^0}\rangle$, {\it i.e.}, 
%%%%%%%%%%%%%%%%%%%%%%%%%%%%%%%%%%%%%%%%%%%%%%%%%%%%%%%%%%%%%% 
\begin{equation}
(\Delta m_K)_{\rm SD} =
{\langle{K^0|H_{\Delta S=2}|\bar K^0}\rangle \over m_K}.   
                                                   \label{eq:m_K-SD}
\end{equation}
%%%%%%%%%%%%%%%%%%%%%%%%%%%%%%%%%%%%%%%%%%%%%%%%%%%%%%%%%%%%%%%%%%
As was seen in {Sec.~II}, it is decomposed into a sum of factorizable 
and non-factorizable parts, 
%%%%%%%%%%%%%%%%%%%%%%%%%%%%%%%%%%%%%%%%%%%%%%%%%%%%%%%%%%%%%%%%
\begin{equation}
(\Delta m_K)_{\rm SD}
= \bigl\{(\Delta m_K)_{\rm SD}\bigr\}_{\rm FA}
+ \bigl\{(\Delta m_K)_{\rm SD}\bigr\}_{\rm NF}.
\end{equation}
%%%%%%%%%%%%%%%%%%%%%%%%%%%%%%%%%%%%%%%%%%%%%%%%%%%%%%%%%%%%%%%%%
The first term on the right-hand-side is estimated by factorizing
$\langle{K^0|H^{BSW}_{\Delta S=2}|\bar K^0}\rangle$, 
%%%%%%%%%%%%%%%%%%%%%%%%%%%%%%%%%%%%%%%%%%%%%%%%%%%%%%%%%%%%%% 
\begin{equation}
\bigl\{(\Delta m_K)_{\rm SD}\bigr\}_{\rm FA}
   =c_{\Delta S=2}\Biggl\{{8\over 3}m_K f_K^2\Biggr\}B_K(\mu), 
                                             \label{eq:m_K-SD-fact}
\end{equation}
%%%%%%%%%%%%%%%%%%%%%%%%%%%%%%%%%%%%%%%%%%%%%%%%%%%%%%%%%%%%%%%%%%
where $B_K(\mu)$ describes the renormalization scale $\mu$ dependence 
of the four-quark operator $O_{\Delta S=2}$. It compensates the 
$\mu$ dependence of the corresponding Wilson coefficient, which is 
given by\cite{NLO}  
%%%%%%%%%%%%%%%%%%%%%%%%%%%%%%%%%%%%%%%%%%%%%%%%%%%%%%%%%%%%%%%%
\begin{equation}
c_{\Delta S=2} \simeq {G_F^2\over 16\pi^2}m_W^2
\Bigl\{\lambda_c^2\eta_1x_c + \lambda_t^2\eta_2S(x_t) 
+ 2\lambda_c\lambda_t\eta_3S(x_c,x_t)
\Bigr\}b(\mu), 
\end{equation}
%%%%%%%%%%%%%%%%%%%%%%%%%%%%%%%%%%%%%%%%%%%%%%%%%%%%%%%%%%%%%%%
where $m_W$ is the weak boson mass and the $\mu$ dependence of 
$\eta_i$, ($i=1,\,2$ and $3$), has been factored out by $b(\mu)$. 
$S(x)$ and $S(x,y)$ are the Inami-Lim functions\cite{Inami-Lim} and 
$x_i=m_i^2/m_W^2$, ($i=c$ and $t$). The so-called $B$ factor, $B_K$, 
is now given by\cite{BBL} 
%%%%%%%%%%%%%%%%%%%%%%%%%%%%%%%%%%%%%%%%%%%%%%%%%%%%%%%%%%%%%%%%%%%%
\begin{equation}
B_K=b(\mu)B_K(\mu).
                                                      \label{eq:B_K}
\end{equation}
%%%%%%%%%%%%%%%%%%%%%%%%%%%%%%%%%%%%%%%%%%%%%%%%%%%%%%%%%%%%%%%%%%

The non-factorizable one is proportional to a matrix element of 
$\tilde O_{\Delta S=2}$ taken between $\langle {K^0}|$ and 
$|{\bar K^0}\rangle$. It can be related to the matrix element of the 
$|\Delta {\bf I}|={3\over 2}$ operator, 
$\tilde O_{\Delta {I}={3/2}}$, in the $|\Delta S|=1$ effective 
weak Hamiltonian, 
%%%%%%%%%%%%%%%%%%%%%%%%%%%%%%%%%%%%%%%%%%%%%
\begin{equation}
\langle{K^0|\tilde O_{\Delta S=2}|\bar K^0}\rangle 
= \sqrt{2}\langle{\pi^0|\tilde O_{\Delta I={3/2}}|\bar K^0}\rangle, 
                                          \label{eq:m_K-SU_f(3)}
\end{equation}
%%%%%%%%%%%%%%%%%%%%%%%%%%%%%%%%%%%%%%%%%%%%%%
in the $SU_f(3)$ symmetry limit~\cite{SU_f(3)} or by using the 
asymptotic $SU_f(3)$ symmetry~\cite{Delta m_K-asymp} which 
implies\cite{asymptotic-symm} a flavor $SU_f(3)$ symmetry of matrix 
elements of operators (like charges, currents, etc.) taken between 
single hadron states with ${\bf 1}$-${\bf 8}$ mixing in the IMF. 
As was seen before, our quark counting leads to the 
$|\Delta {\bf I}|={1\over 2}$ rule for the asymptotic
ground-state-meson matrix elements of $\tilde H_w$. It means that
the right-hand-side of the above equation vanishes so that the
left-hand side also vanishes, i.e., 
%%%%%%%%%%%%%%%%%%%%%%%%%%%%%%%%%%%%%%%%%%%%%%%%%%%%%%%%%%%%%%%%%%%
$\langle{K^0}|\tilde O_{\Delta S=2}|{\bar K^0}\rangle = 0$. 
%%%%%%%%%%%%%%%%%%%%%%%%%%%%%%%%%%%%%%%%%%%%%%%%%%%%%%%%%%%%%%%%%%%
The same quark counting leads directly to the same result as seen
before, and hence we have 
%%%%%%%%%%%%%%%%%%%%%%%%%%%%%%%%%%%%%%%%%%%
\begin{equation}
\{(\Delta m_K)_{\rm SD}\}_{\rm NF}=0.          \label{eq:m_K-SD-NF}
\end{equation}
%%%%%%%%%%%%%%%%%%%%%%%%%%%%%%%%%%%%%%%%%
Therefore we do not need to worry about non-factorizable 
contribution to $(\Delta m_K)_{\rm SD}$. 

The long distance term, $(\Delta m_K)_{\rm LD}$, has been given by 
%%%%%%%%%%%%%%%%%%%%%%%%%%%%%%%%%%%%%%%%%%%%%%%%%%%%
\begin{equation}
 (\Delta m_K)_{\rm LD} = 
\int {dm_n^2 \over 2m_K(m_K^2 - m_n^2)}
        \Bigl\{|\langle{n|H_w|K_L}\rangle |^2 
                  - |\langle{n|H_w|K_S}\rangle |^2 \Bigr\}       
                                             \label{eq:Itzykson}
\end{equation}
%%%%%%%%%%%%%%%%%%%%%%%%%%%%%%%%%%%%%%%%%%%%%%%%%%
in the IMF\cite{Itzykson}, where the states $n$ consist of a complete
set of energy eigen states of hadrons. $(\Delta m_K)_{\rm LD}$ will 
be dominated by contributions of pseudo-scalar meson poles and 
$(\pi\pi)$ intermediate states as discussed in Sec. I, 
%%%%%%%%%%%%%%%%%%%%%%%%%%%%%%%%%%%%%%%%%%%%%%%%%%%%%%%%%%
\begin{equation}
(\Delta m_K)_{\rm LD} 
\simeq (\Delta m_K)_{\rm pole} + (\Delta m_K)_{\pi\pi}. 
                                        \label{eq:Delta m_K-long}
\end{equation}
%%%%%%%%%%%%%%%%%%%%%%%%%%%%%%%%%%%%%%%%%%%%%%%%%%%%%%%%%%
For the $\pi\pi$ continuum contribution, we here take the following 
value\cite{Pennington}, 
%%%%%%%%%%%%%%%%%%%%%%%%%%%%%%%%%%%%%%%%%%%%%%%%%%%%%%
\begin{equation}
{(\Delta m_K)_{\pi\pi} \over \Gamma_{K_S}}
= 0.22 \pm 0.03,                                \label{eq:m_K-pipi}
\end{equation}
%%%%%%%%%%%%%%%%%%%%%%%%%%%%%%%%%%%%%%%%%%%%%%%%%%%%%%%%%%%
which has been obtained by using Omnes-Mushkevili equation and the 
measured $\pi\pi$ phase shifts, where $\Gamma_{K_S}$ is the full 
width of $K_S$. Therefore, we hereafter can concentrate on the pole 
contribution. Since the asymptotic ground-state-meson matrix elements 
of $H_w$ should be  dominated by those of non-factorizable $\tilde
H_w$ as will be seen in {Sec.~VI}, $(\Delta m_K)_{\rm pole}$ is given 
approximately by 
%%%%%%%%%%%%%%%%%%%%%%%%%%%%%%%%%%%%%%%%%%%%%%%%%%%%%%%%%%%
\begin{equation}
(\Delta m_K)_{\rm pole}
\simeq \sum_{P_i}{|\langle{K_L|\tilde H_w|P_i}\rangle|^2 
\over 2m_K(M_K^2 - m_{P_i}^2)}
                                           \label{eq:m_K-pole},
\end{equation}
%%%%%%%%%%%%%%%%%%%%%%%%%%%%%%%%%%%%%%%%%%%%%%%%%%%%%%%%
in the IMF, where $P_i=\pi^0$, $\eta$ and $\eta'$. (The $\iota$ 
contribution has been neglected since it is expected to be not very 
important because of its high mass.) As seen in
Eq.(\ref{eq:m_K-pole}), the pole contribution to the $K_L$-$K_S$ mass 
difference, $(\Delta m_K)_{\rm pole}$, has been described in terms of 
asymptotic ground-state-meson matrix elements of $\tilde H_w$. 

%%%%%%%%%%%%%%%%%%%%%%%%%%%%%%%%%%%%%%%%%%%%%%%%%%%%%%
\section{Two photon decays of $K_L$}

Now we study two photon decays in the IMF for later convenience. The 
amplitude is given by a sum of short distance and long distance ones. 
As mentioned before, it is known~\cite{Gaillard-Lee} that the short 
distance contribution to the $K_L \rightarrow \gamma\gamma$ which is 
given by the triangle diagram is small. Therefore, we neglect it and 
consider only long distance effects. The long distance amplitude is 
again decomposed into a sum of factorizable and non-factorizable
ones. The former is given by factorizing the matrix element of 
$H_w^{BSW}$ taken between $\langle{\gamma\gamma}|$ and 
$|{K_L^0}\rangle$. Under the VMD hypothesis~\cite{VMD}, we obtain 
%%%%%%%%%%%%%%%%%%%%%%%%%%%%%%%%%%%%%%%%%%%%%%%%%%%%%%%%%%%%%%%
\begin{equation}
\langle{\gamma(q)\gamma(k)|H_w^{BSW}|K_L^0(p)}\rangle 
=\sum_{V,V'}\Biggl[{X_V(q^2)\over m_V^2}\Biggr]
             \Biggl[{X_{V'}(k^2)\over m_{V'}^2}\Biggr]
                   \langle{V(q)V'(k)|H_w^{BSW}|K_L^0(p)}\rangle,
\end{equation}
%%%%%%%%%%%%%%%%%%%%%%%%%%%%%%%%%%%%%%%%%%%%%%%%%%%%%%%%%%%%%%
where $V$ and $V'$ denote vector mesons ($\rho^0$, $\omega$ and
$\phi$) which couple to the photon. ${X_V}$ and ${X_{V'}}$ provide 
the corresponding coupling strengths. The above factorizable 
amplitude is strongly suppressed because it is a color suppressed one 
like that for the $K\rightarrow \pi^0\pi^0$ decay. Therefore we can 
neglect safely the factorizable amplitude for the 
$K_L \rightarrow \gamma\gamma$ decay. 

Non-factorizable long distance contributions will be dominated by pole 
amplitudes since contributions of two and more pion intermediate
states are suppressed because of the approximate CP invariance and 
small phase space volume, respectively. However a sum of pseudo 
scalar meson ($P_i=\pi^0,\,\eta,\,\eta'$) pole amplitudes\cite{HPP} 
%%%%%%%%%%%%%%%%%%%%%%%%%%%%%%%%%%%%%%%%%%%%%%%%%%%%%%%%%%%%
\begin{equation}
A_P(K_L \rightarrow \gamma\gamma) = 
\sum_{P_i}{\langle{K_L|\tilde H_w|P_i}\rangle 
A(P_i \rightarrow \gamma\gamma) \over (m_{P_i}^2 - m_K^2)}
                                                 \label{eq:P-pole}
\end{equation}
%%%%%%%%%%%%%%%%%%%%%%%%%%%%%%%%%%%%%%%%%%%%%%
with the usual $\eta$-$\eta'$ mixing angle, 
$\theta_P \simeq -20^\circ$\cite{PDG}, is not 
sufficient\cite{D'Ambrosio-Espriu} to reproduce the observed 
rate\cite{PDG}, 
%%%%%%%%%%%%%%%%%%%%%%%%%%%%%%%%%%%%%%%%%%%%%%%%%%%%%%%
$\Gamma(K_L \rightarrow \gamma\gamma)_{\rm expt} 
= (7.30 \pm 0.33)\times 10^{-12}$ eV. 
%%%%%%%%%%%%%%%%%%%%%%%%%%%%%%%%%%%%%%%%%%%%%%%%%%%%
Therefore we have to take into account some other contributions. 
Although a possible role of the pseudo scalar glue-ball ($\iota$)
through the penguin effect has been considered in Ref.\cite{HPP}, 
it will be not very important because of its high mass and small
rate\cite{PDG},  
%%%%%%%%%%%%%%%%%%%%%%%%%%%%%%%%%%%%%%%%%%%%%%%%%%%%%%%%%%%%%%%%%%%%
$\Gamma(\iota \rightarrow \gamma\gamma)_{\rm expt} < 1.2 $ keV. 
%%%%%%%%%%%%%%%%%%%%%%%%%%%%%%%%%%%%%%%%%%%%%%%%%%%%%%%%%%%%%%%%%%%
Another possible contribution to the $K_L \rightarrow \gamma\gamma$ 
will be the $K^*$ meson pole with the VMD. However there have been 
some arguments against it\cite{Munczek}. These arguments are based on 
the field algebra~\cite{field-algebra} and their weak Hamiltonian 
consists of {\it symmetric} products of left-handed currents and 
transforms like ${\bf 8_s}$ (not ${\bf 8_a}$) of $SU_f(3)$. It is 
much different from the standard model reviewed in Sec.~II. 
Furthermore, the field algebra predicts a finite c-number 
coefficient\cite{VMD} of the Schwinger term\cite{Schwinger}. It 
implies\cite{Dooher} that the cross section, 
%%%%%%%%%%%%%%%%%%%%%%%%%%%%%%%%%%%%%%%%%%%%%%%%%%%%%%%%%%
$\sigma( e^+e^- \rightarrow {\rm hadrons})$, 
%%%%%%%%%%%%%%%%%%%%%%%%%%%%%%%%%%%%%%%%%%%%%%%%%%%%%%%%%%
goes to zero faster than $s^{-2}$ as $s \rightarrow \infty$, where 
$s$ is the total energy in the center of mass system. However it is 
inconsistent with experiment\cite{PDG}. Therefore we should not be 
restricted by such arguments. Since the VMD in the electro-magnetic 
interactions of hadrons can be derived\cite{Bando} independently of 
the field algebra, we now can be free from the arguments in 
Ref.\cite{Munczek} even if we assume the VMD. In this way, we can 
safely take into account the $K^*$ pole contribution in the 
$K_L \rightarrow \gamma\gamma$ decay\cite{Terasaki-FF}. 
Its off-mass-shell amplitude is given by 
%%%%%%%%%%%%%%%%%%%%%%%%%%%%%%%%%%%%%%%%%%%%%%%
\begin{eqnarray}
&& A_{K^*}(K_L \rightarrow \gamma\gamma^*(k^2))
= \sum_{V_i}\sum_{V_j}\sqrt{2}X_{V_i}X_{V_j}
G_{K^0K^{*0}V_i}
\langle{K^{*0}|\tilde H_w|V_j}\rangle_{1}\nonumber\\
&&\hspace{5cm}\times\Bigl\{ 
{1 \over m_{V_i}^2(m_{K^*}^2-k^2)(m_{V_j}^2-k^2)} 
+ {1 \over (m_{V_i}^2 - k^2)m_{K^*}^2m_{V_j}^2}
\Bigr\} 
                                                 \label{eq:K^*-pole}
\end{eqnarray}
%%%%%%%%%%%%%%%%%%%%%%%%%%%%%%%%%%%%%%%%
with $V_i = \rho^0$, $\omega$ and $\phi$. $X_{V_i}$ is the 
photon-vector meson coupling strength. The subscript $1$ of the 
matrix element, $\langle{K^{*0}|\tilde H_w|V_j}\rangle_{1}$, denotes 
the helicity $|\lambda|$ of the vector meson states sandwiching 
$\tilde H_w$ as in Eq.(\ref{eq:para-1}). The $K^*$ pole amplitude for 
the $K_L \rightarrow \gamma\gamma$ decay is simply obtained by 
putting $k^2 = 0$ in the above off-mass-shell amplitude, 
Eq.(\ref{eq:K^*-pole}). 

The pseudo scalar meson pole amplitude, Eq.(\ref{eq:P-pole}), can be 
extrapolated into the off-mass-shell region in the form, 
%%%%%%%%%%%%%%%%%%%%%%%%%%%%%%%%%%%%%%%%%%%%%%%%%%%%%%%%%%%%
\begin{equation}
A_P(K_L \rightarrow \gamma\gamma^*(k^2)) = 
\sum_{P_i}{\langle{K_L|\tilde H_w|P_i}\rangle 
A(P_i \rightarrow \gamma\gamma) 
\over (m_{P_i}^2 - m_K^2)(1 - k^2/\Lambda_{P_i}^2)}.
                                                 \label{eq:P-pole*}
\end{equation}
%%%%%%%%%%%%%%%%%%%%%%%%%%%%%%%%%%%%%%%%%%%%%%
The observed form factors for the $\pi^0$, $\eta$ and 
$\eta' \rightarrow \gamma\gamma^*$ decays are approximately described 
in the form\cite{Landsberg}, $\sim (1 - k^2/\Lambda_P^2)^{-1}$ with 
$\Lambda_P \simeq m_\rho$. For more precise arguments, however, we 
may have to use a more improved result from recent 
measurements\cite{CLEO}, 
%%%%%%%%%%%%%%%%%%%%%%%%%%%%%%%%%%%%%%%%%%%%%%%%%%%%%%%%%%%%%%%%%%%
$\Lambda_\pi = 776 \pm 10 \pm 12 \pm 16 \,\,{\rm MeV}$, 
$\Lambda_\eta = 774 \pm 11 \pm 16 \pm 22  \,\, {\rm MeV}$ 
and $\Lambda_{\eta'} = 859 \pm 9 \pm 18 \pm 20 \,\,{\rm MeV}$. 
%%%%%%%%%%%%%%%%%%%%%%%%%%%%%%%%%%%%%%%%%%%%%%%%%%%%%%%%%%%%%%%%%%%%
In this way, the amplitude and the form factor for the 
$K_L \rightarrow \gamma\gamma^*$ are approximately given by  
%%%%%%%%%%%%%%%%%%%%%%%%%%%%%%%%%%%%%%%%%%%%%%%%%%%%%%%%%%%%%%
\begin{equation}
A(K_L \rightarrow \gamma\gamma^*(k^2)) 
\simeq  A_P(K_L \rightarrow \gamma\gamma^*(k^2)) 
+ A_{K^*}(K_L \rightarrow \gamma\gamma^*(k^2))   
                                                 \label{eq:Dalitz-A}
\end{equation}
%%%%%%%%%%%%%%%%%%%%%%%%%%%%%%%%%%%%%%%%%%%%%%%%%%%%%%%%%%%%%%%%%%%%
and 
%%%%%%%%%%%%%%%%%%%%%%%%%%%%%%%%%%%%%%%%%%%%%%%%%%%%%%%%%%%%%%
\begin{equation}
f(k^2) = {A(K_L \rightarrow \gamma\gamma^*(k^2)) 
\over A(K_L \rightarrow \gamma\gamma)}, 
                                                 \label{eq:Dalitz-FF}
\end{equation}
%%%%%%%%%%%%%%%%%%%%%%%%%%%%%%%%%%%%%%%%%%%%%%%%%%%%%%%%%%%%%%%%%%%%
respectively, where 
%%%%%%%%%%%%%%%%%%%%%%%%%%%%%%%%%%%%%%%%%%%%%%%%%%%
\begin{equation}
A(K_L \rightarrow \gamma\gamma) 
= A(K_L \rightarrow \gamma\gamma^*(k^2 = 0)).    \label{eq:2-gamma}
\end{equation}
%%%%%%%%%%%%%%%%%%%%%%%%%%%%%%%%%%%%%%%%%%%%%%%%%%%
As seen in Eqs.(\ref{eq:Dalitz-A}) and (\ref{eq:Dalitz-FF}) with 
Eqs.(\ref{eq:P-pole}), (\ref{eq:K^*-pole}) and (\ref{eq:P-pole*}), 
the amplitude $A(K_L \rightarrow \gamma\gamma^*)$ and the form factor 
$f(k^2)$ for the $K_L\rightarrow\gamma\gamma^*$ have been written 
approximately in terms of (asymptotic) ground-state-meson matrix 
elements of $\tilde H_w$. 

Since the Dalitz decay of $K_L$ proceeds dominantly as 
%%%%%%%%%%%%%%%%%%%%%%%%%%%%%%%%%%%%%%%%%%%%%%%%%%%%%%%%%%%%%%%% 
$K_L \rightarrow \gamma\gamma^* \rightarrow \gamma \ell^+\ell^-$, 
%%%%%%%%%%%%%%%%%%%%%%%%%%%%%%%%%%%%%%%%%%%%%%%%%%%%%%%%%%%%%%%% 
its branching fraction is given by the following formula\cite{BMS}, 
%%%%%%%%%%%%%%%%%%%%%%%%%%%%%%%%%%%%%%%%%%%%%%%%%%%%%%%%%%%%%%%%%%
\begin{equation}
R_{\gamma\ell^+\ell^-} 
= {\Gamma(K_L \rightarrow \gamma \ell^+\ell^-) 
         \over \Gamma(K_L \rightarrow \gamma \gamma)} 
= \bigl[\Gamma(K_L \rightarrow \gamma \gamma)\bigr]^{-1}
\int_{x_{\rm min}}^{1} dx 
       \Biggl[{d\Gamma(K_L \rightarrow \gamma \ell^+\ell^-)
                                                \over dx}\Biggr] 
                                             \label{eq:Dalitz-rate}
\end{equation}
%%%%%%%%%%%%%%%%%%%%%%%%%%%%%%%%%%%%%%%%%%
with $x=k^2 /m_K^2$, where 
%%%%%%%%%%%%%%%%%%%%%%%%%%%%%%%%%%%%%%%%%%%%%%%%%%%%
\begin{eqnarray}
&& \bigl[\Gamma(K_L \rightarrow \gamma \gamma)\bigr]^{-1}
  {d\Gamma(K_L \rightarrow \gamma \ell^+\ell^-) \over dx} 
                                                       \nonumber\\
&& \hspace{2.5cm}
= \Biggl({2\alpha \over 3\pi}\Biggr){(1 - x)^3 \over x}
   \Biggl[1 + 2\Biggl({m_\ell \over m_K}\Biggr)^2{1 \over x}\Biggr]
\Biggl[1 - 4\Biggl({m_\ell \over m_K}\Biggr)^2{1 \over x}
\Biggr]^{1/2}|f(x)|^2.                          \label{eq:Diff-rate}
\end{eqnarray}
%%%%%%%%%%%%%%%%%%%%%%%%%%%%%%%%%%%%%%%%%%%%%%%%%%%%%%%%%%%%%%%%

%%%%%%%%%%%%%%%%%%%%%%%%%%%%%%%%%%%%%%%%%%%%%%%%%%%%%%%%%%%%%%%
%\newpage
%%%%%%%%%%%%%%%%%%%%%%%%%%%%%%%%%%%%%%%%%%%%%%%%%%%%%%%%%%%%%%%
\section{Comparison with experiments}

Inserting the parameterization of the asymptotic matrix elements of 
$\tilde H_w$ (in Sec.~II and Appendix A) into the general form of the 
non-factorizable long distance amplitudes given in the previous
sections, we can obtain explicit expressions of the amplitudes for 
the $K\rightarrow \pi\pi$ and $K_L\rightarrow\gamma\gamma^{(*)}$ 
decays in addition to the pole contribution, 
$(\Delta m_K)_{\rm pole}$, to the $K_L$-$K_S$ mass 
%%%%%%%%%%%%%%%%%%%%%%%%%%%%%%%%%%%%%%%%%%%%%%%%%%%%%%%%%%
\newpage
%\vspace{0.5cm}
\begin{center}
\begin{quote}
{Table~II. Branching ratios ($\%$) for the $K \rightarrow \pi\pi$
decays. $B_{\rm FA}$, $B_{\rm NF}$ and $B_{\rm TOT}$ are given by the
amplitudes $M_{\rm FA}$, $M_{\rm NF}$ and 
$M_{\rm TOT} = M_{\rm FA} + M_{\rm NF}$, respectively. Values of 
unknown parameters involved in $M_{\rm NF}$ are taken to be 
$\delta_0 = 52^\circ$, $k_a = 0.128$, $k_s = 0.095$ and 
$|\langle{\pi^+|\tilde H_w|K^+}\rangle| = 1.99\times 10^{-7}m_K^2$. 
The observed lifetimes~\cite{PDG}, 
%%%%%%%%%%%%%%%%%%%%%%%%%%%%%%%%%%%%%%%%%%%%%%%%%%%%%%%%%%%%%%%
$\tau(K^\pm)_{\rm expt}=(1.2386\,\pm\,0.0024)\times10^{-8}\,s$ and 
$\tau(K_S^0)_{\rm expt}=(0.8934\,\pm\,0.0008)\times10^{-10}\,s$, 
%%%%%%%%%%%%%%%%%%%%%%%%%%%%%%%%%%%%%%%%%%%%%%%%%%%%%%%%%%%%%%%
have been used as input data. }
\end{quote}
\vspace{0.5cm}

\begin{tabular}
{l|l|l|l|l}
\hline\hline
$\quad{\rm Decay}$
&{$\quad B_{\rm FA}\quad$}
&{$\quad B_{\rm NF}\quad$}
&{$\quad B_{\rm TOT}\quad$}
&{$\quad {\rm Experiment}\quad$}
\\
\hline 
$K_S\, \rightarrow \pi^+\pi^-$
& \quad\,\, 3.2
& \quad 72
& \quad 62
& $\quad 68.61 \pm 0.28$
\\
\hline
$K_S\, \rightarrow \pi^0\,\pi^0$
& \quad\,\, 0.05
& \quad 36
& \quad 35
& $\quad 31.39 \pm 0.28$
\\
\hline
$K^+ \rightarrow \pi^+\pi^0$
& \quad 76
& \quad 22
& \quad 20
& $\quad 21.16 \pm 0.14$
\\
\hline\hline
\end{tabular}

%\end{table}
\end{center}
\vspace{0.5cm}
%%%%%%%%%%%%%%%%%%%%%%%%%%%%%%%%%%%%%%%%%%%%%%%%%%%%%%%%%%%%%%%%%%
difference which 
are summarized in Appendix B. However, they still contain many 
parameters whose values have not been specified. Before we compare 
our result with the measurements, we estimate values of these 
parameters using various experimental data. 

Sizes of the amplitudes $A(P_i \rightarrow \gamma\gamma)$'s in 
Eq.(\ref{eq:P-pole}) can be estimated from the measured decay 
rates\cite{PDG}, 
%%%%%%%%%%%%%%%%%%%%%%%%%%%%%%%%%%%%%%%%%%%%%%%%%
$\Gamma(\pi^0 \rightarrow \gamma\gamma)_{\rm expt} 
= (7.7 \pm 0.6)\,\,{\rm eV}$, 
$\Gamma(\eta \rightarrow \gamma\gamma)_{\rm expt} 
= (0.46 \pm 0.04)\,\,{\rm keV}$ and 
$\Gamma(\eta' \rightarrow \gamma\gamma)_{\rm expt} 
= (4.26 \pm 0.19)\,\,{\rm keV}$. 
%%%%%%%%%%%%%%%%%%%%%%%%%%%%%%%%%%%%%%%%%%%%%%%%%
Their signs are taken to be compatible with the quark model. The 
$V$-$V'$-$P$, ($V,\,V'=K^*,\,\rho,\,\omega$ and $\phi$;
$P= K,\,\pi,\,\eta$ and $\eta'$), coupling constants can be estimated 
from the observed rates for the radiative decays of $K^*$ by using 
$SU_f(3)$ symmetry and the VMD with the $\gamma$-$V$ coupling 
strengths\cite{Terasaki-VMD}, 
%%%%%%%%%%%%%%%%%%%%%%%%%%%%%%%%%%%%%%%%%
$X_{\rho}(0)= 0.033 \pm 0.003$ (GeV)$^2$, 
$X_{\omega}(0)= 0.011 \pm 0.001$ (GeV)$^2$ and 
$X_{\phi}(0)= -0.018 \pm 0.001$ (GeV)$^2$, 
%%%%%%%%%%%%%%%%%%%%%%%%%%%%%%%%%%%%%%%%%%%%%%%%%
estimated from data on photo-productions of vector mesons. 
Although these $\gamma$-$V$coupling strengths can have momentum 
square ($k^2$) dependence, we neglect it in this paper since it is  
mild in the region $k^2 < m_K^2$. From 
%%%%%%%%%%%%%%%%%%%%%%%%%%%%%%%%%%%%%%%%%%%%%%%%%%%%%%%%%%%
$\Gamma(K^{*0} \rightarrow K^0\gamma)_{\rm expt} 
= (0.115 \pm 0.012)$ MeV\cite{PDG}, 
%%%%%%%%%%%%%%%%%%%%%%%%%%%%%%%%%%%%%%%%%%%%%%%%%%%%%%%%%%%%%
we obtain $|G_{K^{*0}K^0\rho^0}| \simeq 0.856$ (GeV)$^{-1}$ and then 
$G_{\omega\pi^0\rho^0} = -2G_{K^{*0}K^0\rho^0}$ using $SU_f(3)$. 
In this way, we can reproduce well the observed rate, 
$\Gamma(\pi^0\rightarrow \gamma\gamma)_{\rm expt}$. 

Size of $\langle{\pi|\tilde H_w|K}\rangle$ can be estimated
phenomenologically by fitting the rates for the $K\rightarrow\pi\pi$ 
decays obtained from total amplitudes (sums of the factorized ones 
in Table~I and the non-factorizable ones in Appendix B) to the 
observed rates. However the latter amplitudes still involve 
many unknown parameters, {\it i.e.}, masses and widths of four-quark 
$[qq][\bar q\bar q]$ and $(qq)(\bar q\bar q)$ mesons, iso-singlet 
$S$-wave $\pi\pi$ phase shift $\delta_0$ at $m_K$, the asymptotic 
ground-state-meson matrix element of $\tilde H_w$, 
$\langle{\pi|\tilde H_w|K}\rangle$, and parameters $k_a$ and $k_s$ 
providing residues of four-quark $[qq][\bar q\bar q]$ and 
$(qq)(\bar q\bar q)$ meson poles as defined in Appendix A. We here 
take the 
predicted values of four-quark meson masses. For their widths, we 
take $\simeq 0.3$ GeV tentatively since they are expected to be 
considerably broad. For the $S$-wave $\pi\pi$ phase shift, we take  
%%%%%%%%%%%%%%%%%%%%%%%%%%%%%%%%%%%%%%%%%%%%%%%%%%%%%%%%%%%%%%%%%%
$\delta_0 \simeq (50-60)^\circ$ at $m_K$\cite{Kamal}. 
%%%%%%%%%%%%%%%%%%%%%%%%%%%%%%%%%%%%%%%%%%%%%%%%%%%%%%%%%%%%%%%%%
Since we do not know, at the present stage, how to estimate values of 
%%%%%%%%%%%%%%%%%%%%%%%%%%%%%%%%%%%%%%%%%%%%%%%%%%%%%%%%%%%%%%
$\langle{\pi|\tilde H_w|K}\rangle$, $k_a$ and $k_s$, 
%%%%%%%%%%%%%%%%%%%%%%%%%%%%%%%%%%%%%%%%%%%%%%%%%%%%%%%%%%%%%
we treat them as adjustable parameters and look for their values to
reproduce the observed rates for the $K\rightarrow\pi\pi$ decays. 
We expect that values of $k_a$ and $k_s$ will be much smaller than
unity since wave function overlapping between the four-quark and the 
ground-state $\{q\bar q\}_0$ meson states is expected to be small and 
that the matrix element, for instance, 
%%%%%%%%%%%%%%%%%%%%%%%%%%%%%%%%%%%%%%%%%%%%
$|\langle{\pi^+|\tilde H_w|K^+}\rangle|$, 
%%%%%%%%%%%%%%%%%%%%%%%%%%%%%%%%%%%%%%%%%%%
will be much larger, because of soft gluon effects, than the 
factorized 
%%%%%%%%%%%%%%%%%%%%%%%%%%%%%%%%%%%%%%%%%%%%%%%%%%%%%%
$|\langle{\pi^+|H_w^{BSW}|K^+}\rangle_{\rm FA}| 
\simeq 0.232m_K^2\times 10^{-7}$. 
%%%%%%%%%%%%%%%%%%%%%%%%%%%%%%%%%%%%%%%%%%%%%%%%%%%%%%
Taking reasonable values of these parameters around $k_a\simeq 0.12$, 
$k_s\simeq 0.10$ and 
%%%%%%%%%%%%%%%%%%%%%%%%%%%%%%%%%%%%%%%%%%%%%%%%%%%%%%%%%%%%%%%%
$|\langle{\pi^+|\tilde H_w|K^+}\rangle|\,
\simeq 2.0m_K^2\times 10^{-7}$,         
%%%%%%%%%%%%%%%%%%%%%%%%%%%%%%%%%%%%%%%%%%%%%%%%%%%%%%%%%%%%%%%%% 
we can reproduce fairly well the observed rates for the 
$K \rightarrow\pi\pi$ decays. In Table~II where branching ratios, 
$B_{\rm FA}$, $B_{\rm NF}$ and $B_{\rm TOT}$, are given by the 
amplitudes, $M_{\rm FA}$, $M_{\rm NF}$ and 
$M_{\rm TOT}=M_{\rm FA}+M_{\rm NF}$, respectively, we show our 
typical result which is obtained by taking the 
following values of parameters, 
%%%%%%%%%%%%%%%%%%%%%%%%%%%%%%%%%%%%%%%%%%%%%%%%%%%%%%%%%%%%%%%%%
$\delta_0 = 52^\circ$, $k_a = 0.128$, $k_s = 0.095$ and 
$|\langle{\pi^+|\tilde H_w|K^+}\rangle| = 1.99m_K^2\times 10^{-7}$, 
%%%%%%%%%%%%%%%%%%%%%%%%%%%%%%%%%%%%%%%%%%%%%%%%%%%%%%%%%%%%%%%%%%%
where the observed lifetimes~\cite{PDG}, 
%%%%%%%%%%%%%%%%%%%%%%%%%%%%%%%%%%%%%%%%%%%%%%%%%%%%%%%%%%%%%%%
$\tau(K^\pm)_{\rm expt}=(1.2386\,\pm\,0.0024)\times10^{-8}\,s$ and 
$\tau(K_S^0)_{\rm expt}=(0.8934\,\pm\,0.0008)\times10^{-10}\,s$, 
%%%%%%%%%%%%%%%%%%%%%%%%%%%%%%%%%%%%%%%%%%%%%%%%%%%%%%%%%%%%%%%
have been used as input data. We will use hereafter the above value of 
$|\langle{\pi^+|\tilde H_w|K^+}\rangle|$ in this paper. As mentioned
before, it is seen that $B_{\rm FA}$ given by the naive factorization 
cannot reproduce the $|\Delta{\bf I}|={1\over 2}$ rule and that the 
non-factorizable contributions play an essential role. As the 
consequence, $B_{\rm TOT}$ can reproduce the observations within 
about 10 $\%$ errors. (Of course, the experimental errors are much 
smaller\cite{PDG}.) 

For $(\Delta m_K)_{\rm SD}$, we have obtained   
%%%%%%%%%%%%%%%%%%%%%%%%%%%%%%%%%%%%%%%%%%%%%%%%%%%%%%%%%%%%%%%%%%%
$\bigl\{(\Delta m_K)_{\rm SD}\bigr\}_{\rm NF}
\propto \langle{K^0|\tilde O_{\Delta S=2}|\bar K^0}\rangle = 0$ 
%%%%%%%%%%%%%%%%%%%%%%%%%%%%%%%%%%%%%%%%%%%%%%%%%%%%%%%%%%%%%%%%%%%%
in Eq.(\ref{eq:m_K-SD-NF}) using [$SU_f(3)$ symmetry and] the quark 
counting in Sec.~II, and hence 
%%%%%%%%%%%%%%%%%%%%%%%%%%%%%%%%%%%%%%%%%%%%%%%%%%%%%%%%%%%%%%%%%%
\begin{equation}
(\Delta m_K)_{\rm SD} 
= \bigl\{(\Delta m_K)_{\rm SD}\bigr\}_{\rm FA}.
\end{equation}
%%%%%%%%%%%%%%%%%%%%%%%%%%%%%%%%%%%%%%%%%%%%%%%%%%%%%%%%%%%%%%%%%%%
Although the value of the coefficient $c_{\Delta S=2}$ at $m_c$ has
been calculated\cite{BBL}, we do not definitely know the value of 
$B_K$. Furthermore, the calculated value of $c_{\Delta S=2}$ (with 
the NLO corrections) contains still large ambiguities. Therefore, we 
will treat $B_K$ as an unknown parameter and study the following 
three cases, (i) $B_K=1.0$, (ii) $B_K=0.75$ and (iii) $B_K=0.50$, 
later since the lattice QCD suggests that 
$B_K \sim 0.6$\cite{B_K-lattice}. When the above NLO 
results\cite{NLO} on $\eta_i$'s are used, top quark contribution to 
$H_{\Delta S=2}$ is not very important as long as real part of the 
$K^0$-$\bar K^0$ mixing amplitude is concerned. Taking 
$\eta_1 \simeq 1.38$\cite{BBL,NLO}, we obtain 
%%%%%%%%%%%%%%%%%%%%%%%%%%%%%%%%%%%%%%%%%%%%%%%%%%%%%%%%%%%%%%%%%%%%
\begin{equation}
{(\Delta m_K)_{\rm SD}\over\Gamma_{K_S}} \simeq 0.49B_K .
\end{equation}
%%%%%%%%%%%%%%%%%%%%%%%%%%%%%%%%%%%%%%%%%%%%%%%%%%%%%%%%%%%%%%%%%% 
Then the pole contribution which has been calculated in Sec.~IV 
should be compared with 
%%%%%%%%%%%%%%%%%%%%%%%%%%%%%%%%%%%%%%%%%%%%%%%%%%%%%%%%%%%%%%%%%%%%%
\begin{eqnarray}
&&{(\Delta m_K)_{\rm pole}\over\Gamma_{K_S}}\,
\simeq \quad\Bigl({\Delta m_K\over\Gamma_{K_S}}\Bigr)_{\rm expt}\quad 
-\quad {(\Delta m_K)_{\pi\pi}\over\Gamma_{K_S}} \quad
-\quad {(\Delta m_K)_{\rm SD}\over\Gamma_{K_S}} 
\nonumber\\
&&\hspace{2cm}\,\,\simeq (0.477 \pm 0.002)\,\,\,
-\,\,\,(0.22 \pm 0.03)\,\, -\quad\,(0.49B_K), 
                                             \label{eq:m_K-pole-num}
\end{eqnarray}
%%%%%%%%%%%%%%%%%%%%%%%%%%%%%%%%%%%%%%%%%%%%%%%%%%%%%%%%%%%%%%%%%%%%%
where the value of $(\Delta m_K)_{\pi\pi}/\Gamma_{K_S}$ has been 
given in Ref.\cite{Pennington} as mentioned before. 

The remaining parameters involved in the $(\Delta m_K)_{\rm pole}$ 
and $\Gamma(K_L\rightarrow \gamma\gamma)$ are 
%%%%%%%%%%%%%%%%%%%%%%%%%%%%%%%%%%%%%%%%%%%%%%%%%%%%%%%%%%%%%%%%%%%
\begin{equation}
\tilde r_0 \quad {\rm and} \quad 
\tilde \alpha_{K^*} 
= {\langle{\rho^+|\tilde H_w|K^{*+}}\rangle_{1}
\over \langle{\pi^+|\tilde H_w|K^+}\rangle},
\end{equation} 
%%%%%%%%%%%%%%%%%%%%%%%%%%%%%%%%%%%%%%%%%%%%%%%%%%%%%%%%%%%%
where $\tilde r_0$ has been defined in Sec.~II as a parameter 
describing the contribution of the penguin relative to $\tilde O_-$ 
in the asymptotic ground-state-meson matrix elements of $\tilde H_w$ 
with the helicity $\lambda=0$. We now search for values of 
$\tilde r_0$ and $\tilde \alpha_{K^*}$ to reproduce 
$(\Delta m_K)_{\rm expt}$ and 
$\Gamma(K_L \rightarrow \gamma\gamma)_{\rm expt}$ 
mentioned before. 
[We have already used $\Gamma(K\rightarrow\pi\pi)_{\rm expt}$'s 
to estimate the size of $\langle{\pi^+|\tilde H_w|K^+}\rangle$.] 
Inserting the parameterization of the asymptotic ground-state-meson 
matrix elements of $\tilde H_w$ in Sec.~II into 
$(\Delta m_K)_{\rm pole}$ in Eq.(\ref{eq:m_K-pole}) and 
$A(K_L \rightarrow \gamma\gamma)$ in Eq.(\ref{eq:2-gamma}) and 
using 
%%%%%%%%%%%%%%%%%%%%%%%%%%%%%%%%%%%%%%%%%%%
$|\langle{\pi^+|\tilde H_w|K^+}\rangle|
\simeq 1.99m_K^2\times 10^{-7}$, 
%%%%%%%%%%%%%%%%%%%%%%%%%%%%%%%%%%%%%%%%%%
we can reproduce the values of 
%%%%%%%%%%%%%%%%%%%%%%%%%%%%%%%%%%%%%%%%%%%%%%%%%%%%%%%%%%%%%%%%%%%%
${(\Delta m_K)_{\rm pole}/\Gamma_{K_S}}$ in 
Eq.(\ref{eq:m_K-pole-num}) 
%%%%%%%%%%%%%%%%%%%%%%%%%%%%%%%%%%%%%%%%%%%%%%%%%%%%%%%%%%%%%%%%%%%%
and 
%%%%%%%%%%%%%%%%%%%%%%%%%%%%%%%%%%%%%%%%%%%%%%%%%%%%%%%%%%%%%%%%%%%%
$\Gamma(K_L \rightarrow \gamma\gamma)_{\rm expt}$ 
%%%%%%%%%%%%%%%%%%%%%%%%%%%%%%%%%%%%%%%%%%%%%%%%%%%%%%%%%%%%%%%%%%%%
(i) for 
%%%%%%%%%%%%%%%%%%%%%%%%%%%%%%%%%%%%%%%%%%%%%%%%%%%%%%%%%%%%%%%
$\tilde r_0 \simeq 0.535$ and $\tilde \alpha_{K^*} \simeq 2.61$ 
%%%%%%%%%%%%%%%%%%%%%%%%%%%%%%%%%%%%%%%%%%%%%%%%%%%%%%%%%%%%%%%%
with $B_K=1.0$, (ii) for 
%%%%%%%%%%%%%%%%%%%%%%%%%%%%%%%%%%%%%%%%%%%%%%%%%%%%%%%%%%%%%%%
$\tilde r_0 \simeq 0.635$ and $\tilde \alpha_{K^*} \simeq 2.34$ 
%%%%%%%%%%%%%%%%%%%%%%%%%%%%%%%%%%%%%%%%%%%%%%%%%%%%%%%%%%%%%%%%
with $B_K=0.75$ and (iii) for 
%%%%%%%%%%%%%%%%%%%%%%%%%%%%%%%%%%%%%%%%%%%%%%%%%%%%%%%%%%%%%%%
$\tilde r_0 \simeq 0.715$ and $\tilde \alpha_{K^*} \simeq 2.15$ 
%%%%%%%%%%%%%%%%%%%%%%%%%%%%%%%%%%%%%%%%%%%%%%%%%%%%%%%%%%%%%%%%
with $B_K=0.50$. All the three cases suggest that
$\tilde r_0 > |c_5/\tilde c_-|$. It means that the ground-state-meson 
matrix element of $O_5$ is actually enhanced relatively to that of 
$\tilde O_-$. It is compatible with $\tilde \alpha_{K^*} > 1$ which 
means that matrix elements of products of left-handed currents taken 
between helicity $\pm 1$ states is enhanced relatively to the ones 
between helicity 0 states. 

Insertion of the parameterization of the asymptotic ground-state-meson 
matrix elements of $\tilde H_w$ with the above values of $\tilde r_0$ 
and $\tilde \alpha_{K^*}$ into Eq.(\ref{eq:Dalitz-FF}) leads to three
different form factors 
for the Dalitz decays of $K_L$. Since the 
difference among the three cases is not very significant, however, we 
show only the result on the form factor for (ii) $B_K=0.75$, 
$\tilde r_0 \simeq 0.635$ and $\tilde \alpha_{K^*} \simeq 2.34$ in 
Fig.~I. For experimental data on the form factor to be compared, 
three different ones have been known, {\it i.e.}, two of them are 
from the $\gamma e^+e^-$ final states\cite{NA31,E845} and the other 
is from the $\gamma\mu^+\mu^-$\cite{E799}. The existing data from 
different types of the final states are not consistent with each 
other near the $\gamma\mu^+\mu^-$ threshold as seen in Fig.~I. 
Our result on the form factor for the 
$K_L\rightarrow \gamma \ell^+\ell^-$ decays seems to prefer to the 
data\cite{NA31,E845} from the $K_L\rightarrow \gamma e^+e^-$ decay. 
Therefore our values of the form factor near the threshold 
of the 
$K_L\rightarrow \gamma \mu^+\mu^-$ decay are considerably larger than 
the data from Ref.\cite{E799}. At higher $x = k^2/m_K^2\,(> 0.4)$,
however, our results are consistent with almost all the data within 
their 
%%%%%%%%%%%%%%%%%%%%%%%%%%%%%%%%%%%%%%%%%%%%%%%%%%%%%%%%%%%%%%%%%%%%%
\vspace{1.0cm}
%\newpage
\begin{center}
\epsfxsize=0.5\hsize
\epsffile{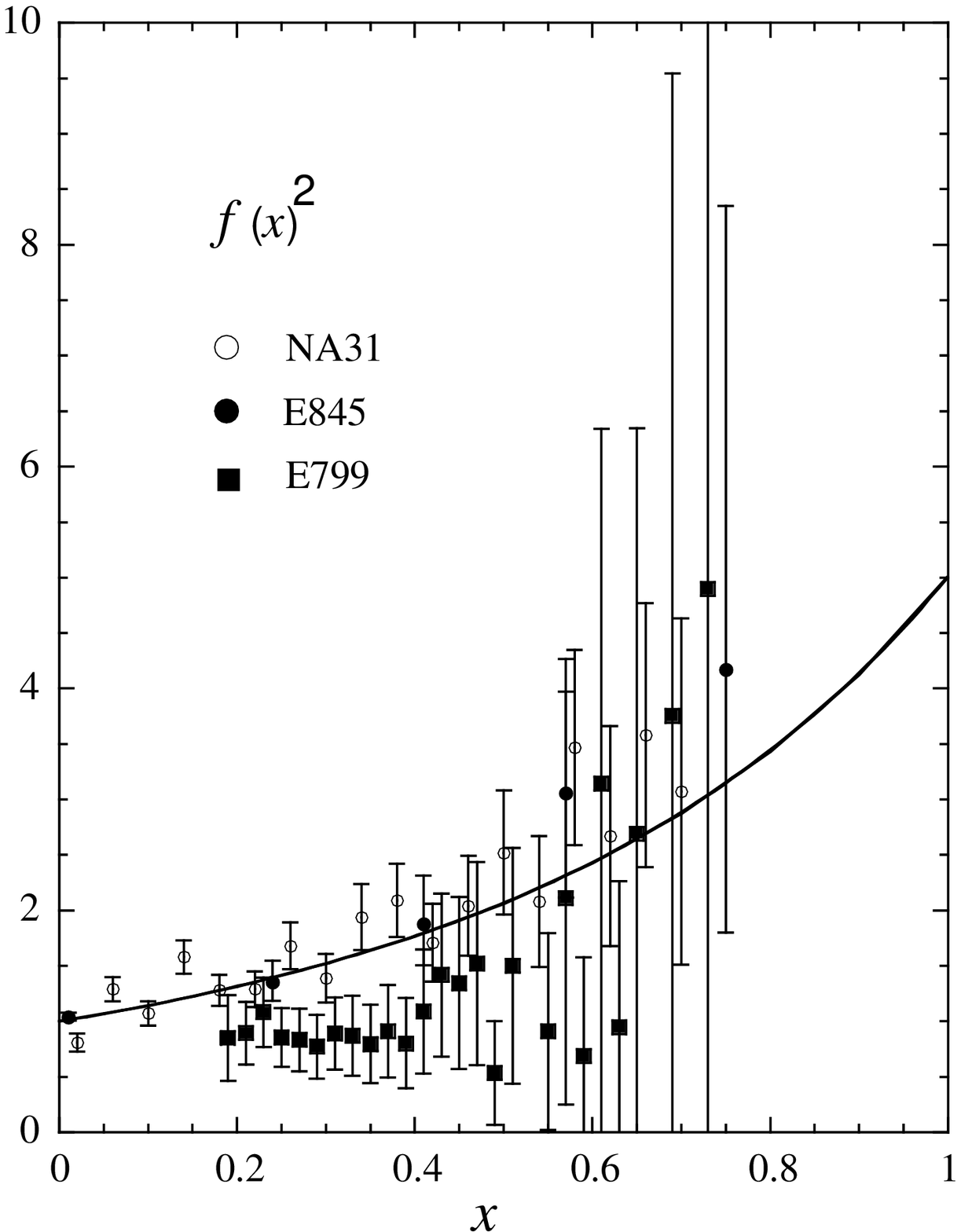}%\\
\vspace{5mm}
\begin{minipage}{130mm}
{Fig.~I. Form factor squared for $B_K=0.75$, $\tilde r_0 = 0.635$ and 
$\tilde \alpha_{K^*} = 2.34$. Circles and bullets are obtained from 
the $\gamma e^+e^-$ final states by E845 at BNL\cite{E845} and NA31 
at CERN\cite{NA31}, respectively, and squares from the 
$\gamma\mu^+\mu^-$ final states by  E799 at FNAL\cite{E799}.}
\vspace{5mm}
\end{minipage}
\end{center}
%%%%%%%%%%%%%%%%%%%%%%%%%%%%%%%%%%%%%%%%%%%%%%%%%%%%%%%%%%%%%%%%%%%%
%%%%%%%%%%%%%%%%%%%%%%%%%%%%%%%%%%%%%%%%%%%%%%%%%%%%%%%%%%
\newpage
%%%%%%%%%%%%%%%%%%%%%%%%%%%%%%%%%%%%%%%%%%%%%%%%%%%%%%%%%%%%%%%%
\begin{center}
\begin{quote}
{Table~III. Branching fractions for the Dalitz decays of $K_L$ for  
(i) $\tilde r_0 = 0.535$ and $\tilde \alpha_{K^*} = 2.61$ 
in the case of $B_K=1.0$, (ii) $\tilde r_0 = 0.635$ and 
$\tilde \alpha_{K^*} = 2.34$ in the case of $B_K=0.75$, and (iii) 
$\tilde r_0 = 0.715$ and $\tilde \alpha_{K^*} = 2.15$ in the case of 
$B_K=0.50$. The data values with ($\ast$), ($\dag$) and ($\ddag$) are 
taken from Refs.\cite{PDG}, \cite{E799} and \cite{NA48},
respectively. 
}
\end{quote}
\vspace{0.5cm}

\begin{tabular}
{c|c|c}
\hline\hline
%\multicolumn{2}{c|}
{}
&$\quad R_{\gamma e^+e^-}(\times 10^{-2})\quad$ 
&$\quad R_{\gamma\mu^+\mu^-}(\times 10^{-4})\quad$
\\
\hline 
(i)\,\, $B_K=1.0$\,\,\,; $\tilde r_0 = 0.535$, $\tilde \alpha_{K^*} = 2.61$ 
&
$1.6$
&$6.5$
\\
\hline
(ii)\, $B_K=0.75$; $\tilde r_0 = 0.635$, $\tilde \alpha_{K^*} = 2.34$ 
&$1.6$
&$6.5$
\\
\hline
(iii) $B_K=0.50$; $\tilde r_0 = 0.715$, $\tilde \alpha_{K^*} = 2.15$ 
&$1.6$
&$6.4$
\\
\hline
%\multicolumn{2}{c|}
{Experiments}
& $(1.6 \pm 0.1)\,\,(\ast)$
& \begin{tabular}{c}
$(5.6 \pm 0.8)\,\,(\dag)$\\
$(5.9 \pm 1.8)\,\,(\ddag)$
\end{tabular}
\\
\hline\hline
\end{tabular}

%\end{table}
\end{center}
%%%%%%%%%%%%%%%%%%%%%%%%%%%%%%%%%%%%%%%%%%%%%%%%%%%%%%%%%%%%%%%%%%
\vspace{0.5cm}
%%%%%%%%%%%%%%%%%%%%%%%%%%%%%%%%%%%%%%%%%%%%%%%%%%%%%%%
large errors. 

Substituting the above results on the form factor for the Dalitz 
decays of $K_L$ into the formula Eq.(\ref{eq:Dalitz-rate}) with 
Eq.(\ref{eq:Diff-rate}), we can calculate their branching
fractions\cite{Erratum} as listed in Table~III. The rate for the 
Dalitz decay of $K_L$ is mainly determined by the values of the 
form factor near the threshold. Therefore, the rate 
$\Gamma(K_L \rightarrow \gamma e^+e^-)$ is not very useful to 
distinguish different theories since its threshold is close to $x=0$ 
where the form factor is normalized to be $f(0) = 1$. Since the 
threshold of the $K_L \rightarrow \gamma \mu^+\mu^-$ decay is 
considerably distant from the normalization point $x = 0$, however, 
we may discriminate different models using this decay when they give
remarkably different values of the form factor around the 
$\gamma\mu^+\mu^-$ threshold. In fact, for example, the BMS 
model\cite{BMS} with $\alpha_{K^*}=-0.28$ which provides the best 
fit to the data on the form factor by NA31\cite{NA31} predicts 
%%%%%%%%%%%%%%%%%%%%%%%%%%%%%%%%%%%%%%%%%%%%%%%%%%%%%%%%%%%%%%%%%%
$R_{\gamma\mu^+\mu^-} \simeq 7.3\times 10^{-4}$
%%%%%%%%%%%%%%%%%%%%%%%%%%%%%%%%%%%%%%%%%%%%%%%%%%%%%%%%%%%%%%%%%
which is substantially larger than our results and the data from 
E799\cite{E799}. The existing theoretical analyses\cite{BMS,Ko} 
involving the above BMS model have been restricted by the argument 
in Ref.\cite{Munczek} which is based on the field algebra and 
therefore, in these theories, the $K^*$ pole 
amplitude has to vanish 
in the $K_L \rightarrow \gamma\gamma$ while it survives in the Dalitz 
decays of $K_L$. In this case, however, it will be 
hard\cite{D'Ambrosio-Espriu} to reproduce the observed 
$\Gamma(K_L\rightarrow \gamma\gamma)$ in consistency with the 
$K\rightarrow\pi\pi$ decays if the usual $\eta$-$\eta'$ mixing angle 
$\theta_P\simeq -20^\circ$ is taken. Additionally, they involve 
implicitly serious problems mentioned before. 

As seen in Table~III, our three results on $R_{\gamma\mu^+\mu^-}$ for 
the above three different values of $B_K$ are not very different from 
each other since the corresponding form factors are not very different 
in the three cases and are a little larger than the data from 
Refs.\cite{E799} but still consistent with the data from 
Ref.\cite{NA48}. It is because our values of the form factor prefer 
to the data from the $K_L\rightarrow \gamma e^+e^-$ 
decay\cite{NA31,E845} but are considerably larger, around the 
$\gamma \mu^+\mu^-$ threshold, than the data by E799\cite{E799} from 
the $K_L\rightarrow \gamma \mu^+\mu^-$ decay. 
%%%%%%%%%%%%%%%%%%%%%%%%%%%%%%%%%%%%%%%%%%%%%%%%%%%%%%%%%%%%%%%%%%
%\newpage
\section{Summary}

We have investigated $K \rightarrow \pi\pi$, $\Delta m_K$, 
$K_L \rightarrow \gamma\gamma$ and the Dalitz decays of $K_L$ 
systematically. Our result has included many parameters. Although
their values have been estimated by using existing experimental data, 
three ($B_K$, $\tilde r_0$ and $\tilde\alpha_{K^*}$) of them have 
still remained unknown. Therefore, in the three cases, (i) $B_K=1.0$
in which $\Delta m_K/\Gamma_{K_S}$ is almost saturated by the short 
distance factorized contribution, (ii) $B_K=0.75$ and 
(iii) $B_K=0.50$ in which about half of the observed value of 
$\Delta m_K/\Gamma_{K_S}$ is occupied by the long distance 
non-factorizable contribution, we have searched for possible values 
of $\tilde r_0$ and $\tilde\alpha_{K^*}$ which reproduce the existing 
{\it data} on $\Delta m_K/\Gamma_{K_S}$ and 
$\Gamma(K_L \rightarrow \gamma\gamma)$ simultaneously. Then, using 
the results, we have calculated the form factor for the Dalitz decays 
of $K_L$ and their decay rates. The three results on the form factor 
have been not very different from each other. On the other hand, the 
existing data on the form  factor from the 
$K_L \rightarrow \gamma e^+e^-$ decay\cite{NA31,E845}  and the 
$K_L \rightarrow \gamma \mu^+\mu^-$ decay\cite{E799} are not 
consistent with each other near the $\gamma \mu^+\mu^-$ threshold.  
Our results on the form factor around the threshold of the 
$K_L \rightarrow \gamma \mu^+\mu^-$ decay seem to be consistent with 
the data from the $\gamma e^+e^-$ final state but a little larger 
than the data from the $\gamma\mu^+\mu^-$ final state. 
The rate for the Dalitz decay is controlled dominantly by the value 
of its form factor near the threshold so that 
$\Gamma(K_L \rightarrow \gamma\mu^+\mu^-)$ will be useful to
distinguish different models in contrast with 
$\Gamma(K_L \rightarrow \gamma e^+e^-)$. However, more theoretical 
and experimental investigations of the Dalitz decays of $K_L$ will be 
needed, since their ambiguities are still large. 
%%%%%%%%%%%%%%%%%%%%%%%%%%%%%%%%%%%%%%%%%%%%%%%%%%%%%%%%%%%%
\vspace{0.5cm}

The author thanks Dr.~K.~E.~Ohl, Dr.~H.~Rohrer, Dr.~D.~Coward and 
Dr.~T.~Nakaya for sending their data values of the form factor for the 
Dalitz decays. He also appreciate Dr.~P.~Singer for arguments against
the $K^*$ pole contribution to the $K_L\rightarrow\gamma\gamma$
decay and Dr. S. Pakvasa and the other members of high energy
physics group, University of Hawaii for discussions, comments and
hospitality during his stay there. 
%%%%%%%%%%%%%%%%%%%%%%%%%%%%%%%%%%%%%%%%%%%%%%%%%%%%%%%%%%%%%%%%
\newpage
\appendix
\section{Asymptotic matrix elements of $\tilde H_w$}

Constraints on asymptotic matrix elements of non-factorizable
four-quark operators $\tilde O_\pm$ have been given in 
Eqs.(\ref{eq:SUM-G}), (\ref{eq:SUM-anti}) and (\ref{eq:SUM-sym}) in 
the text by counting all possible connected quark-line diagrams. 
We here present explicitly the results in the strangeness changing 
case since we have already presented charm changing ones in 
Ref.\cite{TBD-charm-fact}. (Notations of the four-quark mesons have 
also given in Ref.\cite{TBD-charm-fact}.) 

\hfil\break
%%%%%%%%%%%%%%%%%%%%%%%%%%%%%%%%%%%%%%%%%%%%%%%%%%%%%%%%%%%%%%%%%%%%%%
(i) Asymptotic matrix elements of $\tilde H_w$ between ground-state
$\{q\bar q\}_0$ meson states: 

Relations among asymptotic ground-state-meson matrix elements of 
$\tilde O_-$, 
%%%%%%%%%%%%%%%%%%%%%%%%%%%%%%%%%%%%%%%%%%%%%%%%
\begin{equation}
\left\{ 
\begin{array}{l}
\langle{\pi^+|\tilde O_-|K^+}\rangle 
= - \sqrt{2}\langle{\pi^0|\tilde O_-|K^0 }\rangle 
= - \sqrt{2}\langle{\eta_0|\tilde O_-|K^0 }\rangle ,  \\
\langle{\eta_s|\tilde O_-|K^0 }\rangle = 0, 
\end{array} 
\right.
\end{equation}
%%%%%%%%%%%%%%%%%%%%%%%%%%%%%%%%%%%%%%%%%%
and their vector meson counterparts ($K,\,\pi,\,\eta_0$ and 
$\eta_s\rightarrow K^*,\,\rho,\,\omega$ and $\phi$) which are 
compatible with the constraint, Eq.(\ref{eq:SUM-G}), and the 
constraints on the asymptotic matrix elements of $O_5$, 
Eq.(\ref{eq:SUM-O5}), in Sec.~II lead to the 
$|\Delta{\bf I}|={1\over 2}$ rule for the asymptotic 
ground-state-meson matrix elements of $\tilde H_w$, 
%%%%%%%%%%%%%%%%%%%%%%%%%%%%%%%%%%
\begin{equation}
\langle{\pi^+|\tilde H_w|K^+}\rangle 
= - \sqrt{2}\langle{\pi^0|\tilde H_w|K^0 }\rangle, 
\quad   \langle{\rho^+|\tilde H_w|K^{*+}}\rangle 
= - \sqrt{2}\langle{\rho^0|\tilde H_w|K^{*0}}\rangle, 
\end{equation}
%%%%%%%%%%%%%%%%%%%%%%%%%%%%%%%%%%%%%%%%%%%%%%%%%%%%%%%%%%%%%%%%%%%%%
and the parameterizations, Eqs.(\ref{eq:para-0}) and 
(\ref{eq:para-1}), in Sec.~II. 

\hfil\break
(ii) Asymptotic matrix elements of $\tilde H_w$ between 
$\{q\bar q\}_0$ and $[qq][\bar q\bar q]$ meson states: 

Relations among asymptotic matrix elements of $\tilde O_-$ taken
between the ground-state $\{q\bar q\}_0$ and $[qq][\bar q\bar q]$
meson states, 
%%%%%%%%%%%%%%%%%%%%%%%%%%%%%%%%%%%%%
\begin{equation}
\left\{ 
\begin{array}{l}
\langle{\pi^+|\tilde O_-|\hat \kappa^{+}}\rangle 
=-\sqrt{2}\langle{\pi^0|\tilde O_-|\hat \kappa^{0} }\rangle 
=\langle{\sigma|\tilde O_-|K^{0} }\rangle = \cdots, \\
\langle{\sigma^s|\tilde O_-|K^0 }\rangle = 0, 
\end{array} 
\right.
\end{equation}
%%%%%%%%%%%%%%%%%%%%%%%%%%%%%%%%%%%%%
which are compatible with the constraint of Eq.(\ref{eq:SUM-anti})
lead to 
%%%%%%%%%%%%%%%%%%%%%%%%%%%%%%%%%%%%%
\begin{equation}
\langle{\pi^+|\tilde H_w|\hat \kappa^{+}}\rangle 
=-\sqrt{2}\langle{\pi^0|\tilde H_w|\hat \kappa^{0} }\rangle 
=\langle{\sigma|\tilde H_w|K^{0} }\rangle = \cdots 
= { k_a\over 2A_a}\langle{\pi^+\,\,|\tilde H_w|K^+}\rangle,
\end{equation}
%%%%%%%%%%%%%%%%%%%%%%%%%%%%%%%%%%%%%
where $A_a$ is the invariant matrix element of axial charge defined 
by $A_a=-{1\over 2}\langle{\hat\kappa^{+}|A_{\pi^+}|K^0}\rangle$. In
Eq.(A4), we have parameterized the matrix elements using the 
ground-state-meson matrix element of $\tilde H_w$ and a parameter
$k_a$ introduced. In the above, 
asymptotic $SU_f(3)$ symmetry (or nonet symmetry with respect to 
asymptotic matrix elements of charges) has been assumed. 

\hfil\break
(iii) Asymptotic matrix elements of $\tilde H_w$ between 
$\{q\bar q\}_0$ and $(qq)(\bar q\bar q)$ meson states: 

Relations among asymptotic matrix elements of $\tilde O_+$, 
%%%%%%%%%%%%%%%%%%%%%%%%%%%%%%%%%%%%%%%%%%%%%%%%%%%%%%%%%%%%%%%%%%
\begin{eqnarray}
&&\sqrt{{3\over 2}}
             \langle{E_{\pi\pi}^{+}|\tilde O_+|K^+}\rangle 
=-{3\over 2}
          \langle{E_{\pi\pi}^{0}|\tilde O_+|K^0}\rangle 
={3\over \sqrt{2}}\langle{C^{0}|\tilde O_+|K^0}\rangle 
=-{3\over 4}\langle{\pi^{+}|\tilde O_+|E_{\pi K}^+}\rangle  
                                                        \nonumber\\
&&=-{3\over \sqrt{2}}\langle{\pi^{0}|\tilde O_+|E_{\pi K}^0}\rangle 
={\sqrt{3}\over 2}\langle{\pi^{-}|\tilde O_+|E_{\pi K}^-}\rangle 
={3\over \sqrt{2}}\langle{\pi^{+}|\tilde O_+|C_{K}^+}\rangle 
=3\langle{\pi^{0}|\tilde O_+|C_{K}^0}\rangle = \cdots ,
\end{eqnarray}
%%%%%%%%%%%%%%%%%%%%%%%%%%%%%%%%%%%%%%%%%%%%%%%%%%%%%%%%%%%%%%%%%
which are compatible with the constraint of Eq.(\ref{eq:SUM-sym}) 
lead to 
%%%%%%%%%%%%%%%%%%%%%%%%%%%%%%%%%%%%%%%%%%%%%%%%%%%%%%%%%%%%%%%%%%
\begin{eqnarray}
&&\sqrt{{3\over 2}}
             \langle{E_{\pi\pi}^{+}|\tilde H_w|K^+}\rangle 
=-{3\over 2}
          \langle{E_{\pi\pi}^{0}|\tilde H_w|K^0}\rangle 
={3\over \sqrt{2}}\langle{C^{0}|\tilde H_w|K^0}\rangle 
=-{3\over 4}\langle{\pi^{+}|\tilde H_w|E_{\pi K}^+}\rangle  
                                                        \nonumber\\
&&=-{3\over \sqrt{2}}\langle{\pi^{0}|\tilde H_w|E_{\pi K}^0}\rangle 
={\sqrt{3}\over 2}\langle{\pi^{-}|\tilde H_w|E_{\pi K}^-}\rangle 
={3\over \sqrt{2}}\langle{\pi^{+}|\tilde H_w|C_{K}^+}\rangle 
=3\langle{\pi^{0}|\tilde H_w|C_{K}^0}\rangle = \cdots 
                                                        \nonumber\\
&&\hspace{11cm}
= {k_s\over A_s}\langle{\pi^0|\tilde H_w|K^0}\rangle, 
\end{eqnarray}
%%%%%%%%%%%%%%%%%%%%%%%%%%%%%%%%%%%%%%%%%%%%%%%%%%%%%%%%%%%%%%%%%
where $A_s$ is the invariant matrix element of axial charge defined 
by $A_s=\langle{C_K^{+}|A_{\pi^+}|K^0}\rangle$. The above equations
imply that the asymptotic matrix elements of $\tilde H_w$ between 
$\{q\bar q\}_0$ and $(qq)(\bar q\bar q)$ meson states can violate the 
$|\Delta{\bf I}|={1\over 2}$ rule. In Eq.(A6), we have parameterized 
the matrix elements using the ground-state-meson matrix element of 
$\tilde H_w$ and a parameter $k_s$ introduced. In the above, 
asymptotic $SU_f(3)$ symmetry (or nonet symmetry with respect 
to asymptotic matrix elements of charges) has been assumed. 

%%%%%%%%%%%%%%%%%%%%%%%%%%%%%%%%%
%\newpage
\section{Non-factorizable amplitudes}

We here list approximate non-factorizable amplitudes for the 
$K\rightarrow\pi\pi$ decays, pole contribution to the $K_L$-$K_S$ mass 
difference and long distance amplitude for the 
$K_L \rightarrow \gamma\gamma^{(*)}$. 

\hfil\break
(i) Non-factorizable amplitudes for the $K \rightarrow \pi\pi$ decays:

Inserting the constraints on asymptotic matrix elements of 
$\tilde H_w$ in Appendix A into Eq.(\ref{eq:hard pion}) with 
Eqs.(\ref{eq:ETC}) and (\ref{eq:SURF}), we obtain the following 
non-factorizable amplitudes for the $K\rightarrow \pi\pi$ decays, 
%%%%%%%%%%%%%%%%%%%%%%%%%%%%%%%%%%%%%%%%%%%%%%%%%%%%%%%%%%%%%%%
\begin{eqnarray}
&&M_{\rm NF}(K_S^0 \rightarrow \pi^+\pi^-)  \nonumber \\
&&\hspace{2cm}
\simeq -{i\over f_\pi}\langle{\pi^+|\tilde H_w|K^+}\rangle
\Biggl\{
e^{i\delta_0(\pi\pi)}
-\Bigl[2\Bigl({m_K^2-m_\pi^2 \over m_{\hat \sigma}^2-m_K^2}\Bigr)
+\Bigl({m_K^2-m_\pi^2 \over m_{\hat \kappa}^2-m_\pi^2}\Bigr)\Bigr]k_a
                                                         \nonumber\\
&&{\hspace{7cm}}
+\Bigl[2\Bigl({m_K^2-m_\pi^2 \over m_{E_{\pi\pi}}^2-m_K^2}\Bigr)
+\Bigl(
{m_K^2-m_\pi^2 \over m_{E_{\pi K}}^2-m_\pi^2}\Bigr)\Bigr]k_s
\Biggr\},
\end{eqnarray}
%%%%%%%%%%%%%%%%%%%%%%%%%%%%%%%%%%%%%%%%%%%%%%%%%%%%%%%%%%%%%%%
%\newpage
\begin{eqnarray}
&&M_{\rm NF}(K_S^0 \rightarrow \pi^0\pi^0)    \nonumber\\
&&\hspace{1.3cm}
\simeq -{i\over f_\pi}\langle{\pi^+|\tilde H_w|K^+}\rangle
\sqrt{1\over 2}\Biggl\{
e^{i\delta_0(\pi\pi)}
-\Bigl[2\Bigl({m_K^2-m_\pi^2 \over m_{\hat \sigma}^2-m_K^2}\Bigr)
+\Bigl({m_K^2-m_\pi^2 \over m_{\hat \kappa}^2-m_\pi^2}
                                     \Bigr)\Bigr]k_a \nonumber\\
&&{\hspace{7cm}}
-\Bigl[2\Bigl({m_K^2-m_\pi^2 \over m_{E_{\pi\pi}}^2-m_K^2}\Bigr)
+\Bigl(
{m_K^2-m_\pi^2 \over m_{E_{\pi K}}^2-m_\pi^2}\Bigr)\Bigr]k_s
\Biggr\},
\end{eqnarray}
%%%%%%%%%%%%%%%%%%%%%%%%%%%%%%%%%%%%%%%%%%%%%%%%%%%%%%%%%%%%%%%
\begin{equation}
M_{\rm NF}(K^+ \rightarrow \pi^+\pi^0)   
%&&{\hspace{3.5cm}}
\simeq -{i\over f_\pi}\langle{\pi^+|\tilde H_w|K^+}\rangle
\Biggl\{
\Bigl[2\Bigl({m_K^2-m_\pi^2 \over m_{E_{\pi\pi}}^2-m_K^2}\Bigr)
+\Bigl(
{m_K^2-m_\pi^2 \over m_{E_{\pi K}}^2-m_\pi^2}\Bigr)\Bigr]k_s
\Biggr\},
\end{equation}
%%%%%%%%%%%%%%%%%%%%%%%%%%%%%%%%%%%%%%%%%%%%%%%%%%%%%%%%%%%%%%%%%
where pole contributions of vector mesons, orbitally and radially 
excited mesons have been neglected and the mass 
relations\cite{Jaffe}, 
%%%%%%%%%%%%%%%%%%%%%%%%%%%%%%%%%%%%%%%%%%%%%%%%%
$m_{E_{\pi\pi}}=m_C$ and $m_{E_{\pi K}}=m_{C_K}$, 
%%%%%%%%%%%%%%%%%%%%%%%%%%%%%%%%%%%%%%%%%%%%%%%%%
have been used. 

\hfil\break
(ii) Pole contribution to the $K_L$-$K_S$ mass difference: 

Inserting the parameterization of the asymptotic ground-state-meson 
matrix elements of $\tilde H_w$ in Eqs.(\ref{eq:para-0}) and 
(\ref{eq:para-1}) into Eq.(\ref{eq:m_K-pole}), we obtain the following 
pole contribution to $\Delta m_K$, 
%%%%%%%%%%%%%%%%%%%%%%%%%%%%%%%%%%%%%%%%%%%%%%%%%%%%%%%%%%%%%%%
\begin{equation}
{(\Delta M_K)_{\rm pole}\over \Gamma_{K_S}}  
 \simeq {|\langle{\pi^0|\tilde H_w|\bar K^0}\rangle|^2 
\over m_K\Gamma_{K_S}(m_K^2 - m_\pi^2)}
\Biggl\{1 + 
\Biggl({m_K^2 - m_\pi^2 \over m_K^2 - m_\eta^2}\Biggr)P_\eta^2 
+ \Biggl({m_K^2 - m_\pi^2 \over m_K^2 - m_{\eta'}^2}\Biggr)P_{\eta'}^2
\Biggr\},
\end{equation}
%%%%%%%%%%%%%%%%%%%%%%%%%%%%%%%%%%%%%%%%%%%%%%%%%%%%%%%%%%%%%%%%%
where 
%%%%%%%%%%%%%%%%%%%%%%%%%%%%%%%%%%%%%%%%%%%%%%%%%%%%%%%%%%%%%%%
\begin{equation}
P_\eta 
= \Bigl({1 - \tilde r_0 \over 1 + \tilde r_0}\Bigr)a_\eta^0 
- \Bigl({\sqrt{2}\tilde r_0 \over 1 + \tilde r_0}\Bigr)a_\eta^s
\quad{\rm and}\quad 
P_{\eta'} 
= \Bigl({1 - \tilde r_0 \over 1 + \tilde r_0}\Bigr)a_{\eta'}^0 
- \Bigl({\sqrt{2}\tilde r_0 \over 1 + \tilde r_0}\Bigr)a_{\eta'}^s. 
\end{equation}
%%%%%%%%%%%%%%%%%%%%%%%%%%%%%%%%%%%%%%%%%%%%%%%%%%%%%%%%%%%%%
$a_i^0$ and $a_i^s$, ($i=\eta$ and $\eta'$), are the $\eta$-$\eta'$ 
mixing parameters whose explicit expression is given in 
Eqs.(\ref{eq:eta}) in the text.  

\hfil\break
(iii) Non-factorizable long distance amplitude for the
$K_L\rightarrow\gamma\gamma^{*}$ decay:

Inserting the parameterization of the asymptotic ground-state-meson 
matrix elements of $\tilde H_w$ in Eqs.(\ref{eq:para-0}) and 
(\ref{eq:para-1}) into Eq.(\ref{eq:Dalitz-A}) with 
Eqs.(\ref{eq:K^*-pole}) and (\ref{eq:P-pole*}) in the text, we obtain 
the following long distance amplitude for the
$K_L\rightarrow\gamma\gamma^*$ decay, 
%%%%%%%%%%%%%%%%%%%%%%%%%%%%%%%%%%%%%%%%%%%%%%%%%%%%%%%%%%%%%%
\begin{equation}
A(K_L \rightarrow \gamma\gamma^*(k^2)) 
\simeq  A_P(K_L \rightarrow \gamma\gamma^*(k^2)) 
+ A_{K^*}(K_L \rightarrow \gamma\gamma^*(k^2)),   
                                                 \label{eq:Dalitz}
\end{equation}
%%%%%%%%%%%%%%%%%%%%%%%%%%%%%%%%%%%%%%%%%%%%%%%%%%%%%%%%%%%%%%%%%%%%
where 
%%%%%%%%%%%%%%%%%%%%%%%%%%%%%%%%%%%%%%%%%%%%%%%%%%%%%%%%%%%%%%%%
\begin{eqnarray}
&& A_P(K_L\rightarrow\gamma\gamma^* )                
 \simeq {\langle{\pi^0|\tilde H_w|\bar K_L^0}\rangle 
\over (m_K^2 - m_\pi^2)}                  
%\hspace{0.2cm}
\times\Biggl\{{A(\pi^0\rightarrow\gamma\gamma) 
 \over 1 - 
\Bigl(\displaystyle{m_K\over \Lambda_{\pi^0}}\Bigr)^2 x } 
 \nonumber \\
&&+ \Biggl[{A(\eta\rightarrow\gamma\gamma) \over 
{1 - \Bigl(\displaystyle{m_K\over\Lambda_{\eta}}\Bigr)^2 x }}\Biggr]
\Biggl({m_K^2 - m_\pi^2 \over m_K^2 - m_\eta^2}\Biggr)P_\eta    
+ \Biggl[{A(\eta'\rightarrow\gamma\gamma) 
      \over 
{1 - \Bigl(\displaystyle{m_K\over\Lambda_{\eta'}}\Bigr)^2 x} }\Biggr]
\Biggl({m_K^2 - m_\pi^2 \over m_K^2 - m_{\eta'}^2}\Biggr)P_{\eta'}
\Biggr\}                                                     
\end{eqnarray}
%%%%%%%%%%%%%%%%%%%%%%%%%%%%%%%%%%%%%%%%%%%%%%%%%%%%%%%%%%%%%%%%%%%%%%%
and 
%%%%%%%%%%%%%%%%%%%%%%%%%%%%%%%%%%%%%%%%%%%%%%%%%%%%%%%%%%%%%%%%%%%%%%%
%\newpage
\begin{eqnarray}
&& A_V(K_L\rightarrow\gamma\gamma^* )                
 \simeq \sqrt{2}{\langle{\rho^0|\tilde H_w|\bar K^{*0}}\rangle 
                                              G_{K^{*0}K^0\rho^0}
\over (m_{K^*}^2 - k^2)(m_{\rho}^2 - k^2)}   
\Bigl[{X_\rho(0) \over m_\rho^2}\Bigr]X_\rho(k^2)    
 F_{KK^*V}(0)F_{K^*V}(k^2)                 \nonumber\\
%\vspace{2mm}
&&\vspace{2mm}\hspace{3cm} +   
\sqrt{2}{\langle{\rho^0|\tilde H_w|\bar K^{*0}}\rangle 
                                              G_{K^{*0}K^0\rho^0}
\over m_{K^*}^2(m_{\rho}^2 - k^2)}   
\Bigl[{X_\rho(0) \over m_\rho^2}\Bigr]X_\rho(k^2)    
 F_{KK^*V}(k^2)F_{K^*V}(0)     
\end{eqnarray}
%%%%%%%%%%%%%%%%%%%%%%%%%%%%%%%%%%%%%%%%%%%%%%%%%%%%%%%%%%%%%%%%%
with 
%%%%%%%%%%%%%%%%%%%%%%%%%%%%%%%%%%%%%%%%%%%%%%%%%%%%%%%%%%%%%%%
\begin{equation}
F_{KK^*V}(k^2) 
= 1 +  {G_{K^{*0}K^0\omega}\over G_{K^{*0}K^0\rho^0}}
{X_\omega(k^2)\over X_\rho(k^2)}
\Biggl({m_\rho^2 - k^2 \over m_\omega^2 - k^2}\Biggr)
+ {G_{K^{*0}K^0\phi}\over G_{K^{*0}K^0\rho^0}}
{X_\phi(k^2)\over X_\rho(k^2)}
\Biggl({m_\rho^2 - k^2 \over m_\phi^2 - k^2}\Biggr)
\end{equation}
%%%%%%%%%%%%%%%%%%%%%%%%%%%%%%%%%%%%%%%%%%%%%%%%%%%%%%%%%%%%%%%
and 
%%%%%%%%%%%%%%%%%%%%%%%%%%%%%%%%%%%%%%%%%%%%%%%%%%%%%%%%%%%%%%%%%%
\begin{equation}
F_{K^*V}(k^2)
=1 +  \Biggl({1 - \tilde r_1 \over 1 + \tilde r_1}\Biggr)
{X_\omega(k^2)\over X_\rho(k^2)}
\Biggl({m_\rho^2 - k^2 \over m_\omega^2 - k^2}\Biggr)
- \Biggl({\sqrt{2}\tilde r_1 \over 1 + \tilde r_1}\Biggr)
{X_\phi(k^2)\over X_\rho(k^2)}
\Biggl({m_\rho^2 - k^2 \over m_\phi^2 - k^2}\Biggr). 
\end{equation}
%%%%%%%%%%%%%%%%%%%%%%%%%%%%%%%%%%%%%%%%%%%%%%%%%%%%%%%%%%%%

%%%%%%%%%%%%%%%%%%%%%%%%%%%%%%%%%

%%%%%%%%%%%%%%%%%%%%%%%%%%%%%%%%%%%%%%%%%%%%%%%%%%%%%%%%%%%%%%%%%%%%

\begin{thebibliography}{99}
\bibitem{QCD-corr} Although the next-leading-order corrections have 
already been calculated in Ref.\cite{BBL}, the result seems to be 
sensitive to a choice of renormalization scheme in contrast with the 
leading order corrections. We here take the latter since we do not
need very precise values of $a_1$ and $a_2$.
\bibitem{quark-counting} We have considered, for long time, 
constraints on asymptotic matrix elements of the total weak
Hamiltonian $H_w$ which includes both of the factorizable and 
non-factorizable ones since we did not have the idea that the 
Hamiltonian should be decomposed into two parts as discussed in 
Sec.~II. However, it has been improved recently in 
Ref.\cite{TBD-charm-fact}.
\bibitem{asymptotic-symm} A comprehensive review on the asymptotic
flavor symmetry and its fruitful results has been provided in
Ref.\cite{suppl}.
\bibitem{Erratum} Our previous results in Ref.\cite{Terasaki-FF} on 
the branching fraction, $R_{\gamma e^+e^-}$,  were too small since 
the singular behavior of its kinematical factors near the threshold 
was not sufficiently taken into account. They have been improved in 
Table~II.
\end{thebibliography}
\end{document}